%% file: 00-main.tex
\documentclass[manuscript]{acmart}
\usepackage{comment}
\usepackage{csquotes}
\usepackage{tabularx}
\usepackage{enumitem}
\usepackage{booktabs}
\usepackage{array}
\usepackage[dvipsnames]{xcolor}
\usepackage{soul}
\setlist[itemize]{noitemsep}
\usepackage{listings}
\usepackage[most]{tcolorbox}

\newcommand{\eg}{\textit{e.g., }}
\newcommand{\ie}{\textit{i.e. }}

\newcommand{\etal}{\textit{et al. }}

\newtcolorbox{implication}[2][]{
    colback=white,           % Background color of the body
    colframe=black,          % Border color
    colbacktitle=black,      % Title bar background
    coltitle=white,          % Title text color
    fonttitle=\bfseries\sffamily,
    sharp corners=false,     % Rounded corners
    arc=5pt,                 % Radius of the curves
    left=10pt, right=10pt,   % Padding
    top=10pt, bottom=10pt,
    title={#2},              % The title text
    enhanced,                % Allows for advanced skinning
    attach boxed title to top left={yshift=-2mm, xshift=2mm},
    #1                       % Placeholder for extra options
}

\newlength\myindent
\renewcommand\footnotetextcopyrightpermission[1]{} % removes footnote with conference information in first column

\AtBeginDocument{%
  \providecommand\BibTeX{{%
    \normalfont B\kern-0.5em{\scshape i\kern-0.25em b}\kern-0.8em\TeX}}}

\setcopyright{acmcopyright}
\copyrightyear{2025}
\acmYear{2025}
\acmDOI{XXXXXXX.XXXXXXX}

\begin{document}
\input{10-template/commands}

\title[Middleware for Feed Recommendation in Practice]{Middleware for Feed Recommendation in Practice: How Feed Creators Build, Maintain, and Sustain Custom Feeds on Bluesky}

%Understanding Middleware-Driven Social Media Feed Recommendation through Bluesky Feed Creators

% \renewcommand{\shortauthors}{Anonymous authors}

\author{Tony Zhou}
\affiliation{%
  \institution{University of Washington}
  \city{Seattle}
  \state{WA}
  \country{USA}}
\email{tyzhou05@uw.edu}

\author{Leijie Wang}
\affiliation{%
  \institution{University of Washington}
  \city{Seattle}
  \state{WA}
  \country{USA}}
\email{leijiew@cs.washington.edu}

\author{Amy X. Zhang}
\affiliation{%
  \institution{University of Washington}
  \city{Seattle}
  \state{WA}
  \country{USA}}
\email{axz@cs.uw.edu}

\begin{abstract}
\input{sections/00-abstract}
\end{abstract}




% \begin{teaserfigure}
%   \includegraphics[width=\textwidth]{figures/test-teaser.pdf}
%   \caption{Our taxonomy of custom feed roles and characteristics on Bluesky. We characterize custom feed (i) logic, (ii) feed creator, and (iii) subscribers}
%   \label{fig:keyscreen}
%   \Description{.}
% \end{teaserfigure}

\maketitle
\input{sections/01-introduction}
\input{sections/02-related-work}
\input{sections/03-method}
\input{sections/04-findings}
\input{sections/06-discussion}
\input{sections/07-limitations}
\input{sections/08-conclusion}

\bibliographystyle{ACM-Reference-Format}
\bibliography{99-bib}

\newpage
\appendix
\input{sections/appendix}

\end{document}

%% file: 10-template/commands.tex
% Editors
\newcommand{\gromit}[1]{$<$\textcolor{red}{Gromit: #1}$>$}
\newcommand{\donghoon}[1]{$<$\textcolor{orange}{Donghoon: #1}$>$}

% TODO's
\newcommand\todoit[1]{{\color{red}\{TODO: \textit{#1}\}}}
\newcommand\todocite{{\color{red}{CITE}}}

% Highlights colors.
\definecolor{lightblue}{RGB}{212, 235, 255}
\definecolor{orange}{RGB}{255, 105, 0}
\definecolor{lightgreen}{RGB}{177, 231, 171}
\definecolor{lightyellow}{RGB}{255, 255, 148}

% Tables
\newcolumntype{P}[1]{>{\centering\arraybackslash}p{#1}}
\newcolumntype{L}[1]{>{\raggedright\let\newline\\\arraybackslash\hspace{0pt}}m{#1}}
\newcolumntype{R}[1]{>{\raggedleft\arraybackslash}p{#1}}
\newcommand\tworows[1]{\multirow{2}{*}{\shortstack[l]{#1}}}
\newcommand\tworowsc[1]{\multirow{2}{*}{\shortstack[c]{#1}}}
\newcommand\threerows[1]{\multirow{3}{*}{\shortstack[l]{#1}}}

%% file: sections/00-abstract.tex
Scholars have long proposed third-party \textit{middleware} as an alternative to centralized algorithmic feeds: feeds built and distributed by independent \textit{feed creators}. This vision saw no large-scale instantiation until Bluesky, a decentralized microblogging platform, introduced \textit{custom feeds} in 2023. Although central to the middleware ecosystem, we know little about how feed creators understand their role, build feeds, and sustain them. Through interviews with $n = 26$ feed creators and third-party developers of feed-building tools, and analysis of $n = 88,302$ custom feeds, we identify two creator orientations---\textit{utility-providing} and \textit{community-building}. Additionally, creators struggle to maintain feeds that fully realize middleware ideals: they lack granular interaction data, receive little feedback, and lack technical expertise to act on either.
Finally, creators sustain their feeds as unpaid hobbyists with little platform support and are divided on whether to monetize beyond covering costs.
We conclude with design and policy implications for strengthening the middleware feed ecosystem.

%% file: sections/01-introduction.tex
\input{figures/main-figure}

\section{Introduction}

Social media platforms increasingly organize users’ online experiences through centralized, top-down algorithmic feeds. They determine what content is surfaced, prioritized, or suppressed, leaving users little ability to understand or meaningfully steer the logic behind what they see~\cite{kleinberg2024challenge, alvarado2018towards, ananny2018seeing, swart2021experiencing}. In response, scholars have proposed an alternative vision of social media: \textit{middleware}, an open, composable layer where users can choose among third-party services for core components of online experiences such as recommendation and moderation~\cite{fukuyama2020middleware, keller2021future}. This vision posits that opening recommendation and moderation to a competitive ecosystem of third-party providers would decentralize platform power over users' online experiences~\cite{hogg2024shaping,jhaver2023decentralizing}.
However, realizing these benefits also shifts challenges of governance, accountability, and sustainability from platform providers down to middleware operators~\cite{masnickprotocolsnotplatforms, keller2021future}.

While many have written about this vision, only in the last few years has it taken shape at scale. On Bluesky, a decentralized microblogging platform built around the goal of enabling user-driven ``algorithmic choice''~\cite{bluesky-algorithmic-choice}, platform developers instantiated a feature in 2023 called \textit{custom feeds}, which allows users to freely create, govern, and distribute feed recommendation algorithms on the platform~\cite{kleppmann2024bluesky}. 
As this custom feed ecosystem grew, several new user roles arose to support custom feed infrastructure, including \textit{feed creators} who author and maintain feeds, \textit{feed contributors} whose content populates various feeds, \textit{feed consumers} who subscribe to and browse feeds, and \textit{third-party developers} who build tooling that supports feed creation and hosting. As of writing, Bluesky has grown to over 45 million total users and more than 100 thousand custom feeds~\cite{jaz-stats, blue-facts}.
We thus argue that the custom feed ecosystem on Bluesky represents the first large-scale instantiation of third-party middleware for social media feed recommendation. 

Despite being central to this middleware ecosystem, feed creators remain understudied beyond a few quantitative characterizations of the broader ecosystem~\cite{quelle2025bluesky, failla2024m, balduf2024looking}. These analyses find that feeds are created and sustained by a relatively small population of creators, making it all the more important to understand their motivations, work, and challenges. While much research has studied community moderators on platforms such as Reddit or Discord~\cite{seering2022metaphors, weld2025perceptions}, feed creators occupy a fundamentally different space with distinct interactions and user expectations from those community settings. As such, we ask the following research questions:

\begin{itemize}
    % \item \edit{\textbf{RQ1:} What role do feed creators play in the custom feed ecosystem, and how do their efforts manifest broadly across the ecosystem?}
    \item {\textbf{RQ1:} How do feed creators understand their role, and how do these orientations manifest in user activity across the custom feed ecosystem?}
    \item {\textbf{RQ2:} How do feed creators and third-party developers build custom feeds and what tradeoffs shape their choices?}
    \item {\textbf{RQ3:} How are feed creators and third party developers maintaining and sustaining the custom feed ecosystem over time?}
\end{itemize}

% \begin{itemize}
%     \item {\textbf{RQ1:} Why do feed creators build custom feeds, and what kinds of content and algorithms do they serve?}
%     \item {\textbf{RQ2:} How are feed creators and third party developers maintaining and sustaining the custom feed ecosystem on Bluesky?}
%     \item {\textbf{RQ3:} What challenges and opportunities exist for the custom feed ecosystem, and the future of third party middleware for social media feed recommendation?}
% \end{itemize}

To answer these research questions, we interviewed $n = 24$ Bluesky feed creators and $n = 2$ third-party developers of feed-related tools and conducted a large-scale analysis of $n = 88,302$ custom feeds. We identify two distinct orientations of feed creators---\textit{utility-providing} and \textit{community-building}---that reflect distinct goals, maintenance practices, and relationships between feed creators and consumers. Despite the ecosystem's open architecture, we discern dynamics in power: Bluesky retains privileged access to interaction data that custom feeds cannot yet match, and only a handful of creators hold the technical expertise to build powerful feeds beyond third-party tooling. At the same time, the ecosystem is sustained entirely through volunteer hobbyist labor, with creators and third party developers absorbing both the time costs of maintenance (\eg promotion) and the financial costs of sustainment (\eg server hosting). We conclude by offering a framework of power in the custom feed ecosystem and design implications for improving the current custom feed ecosystem and future middleware-based social media systems.

% We further discuss how recommendation power, labor, and accountability are redistributed within decentralized feed ecosystems. We conclude by offering implications for designing future middleware-based social media systems.

%% file: figures/main-figure.tex
\begin{figure*}
  \includegraphics[width=\textwidth]{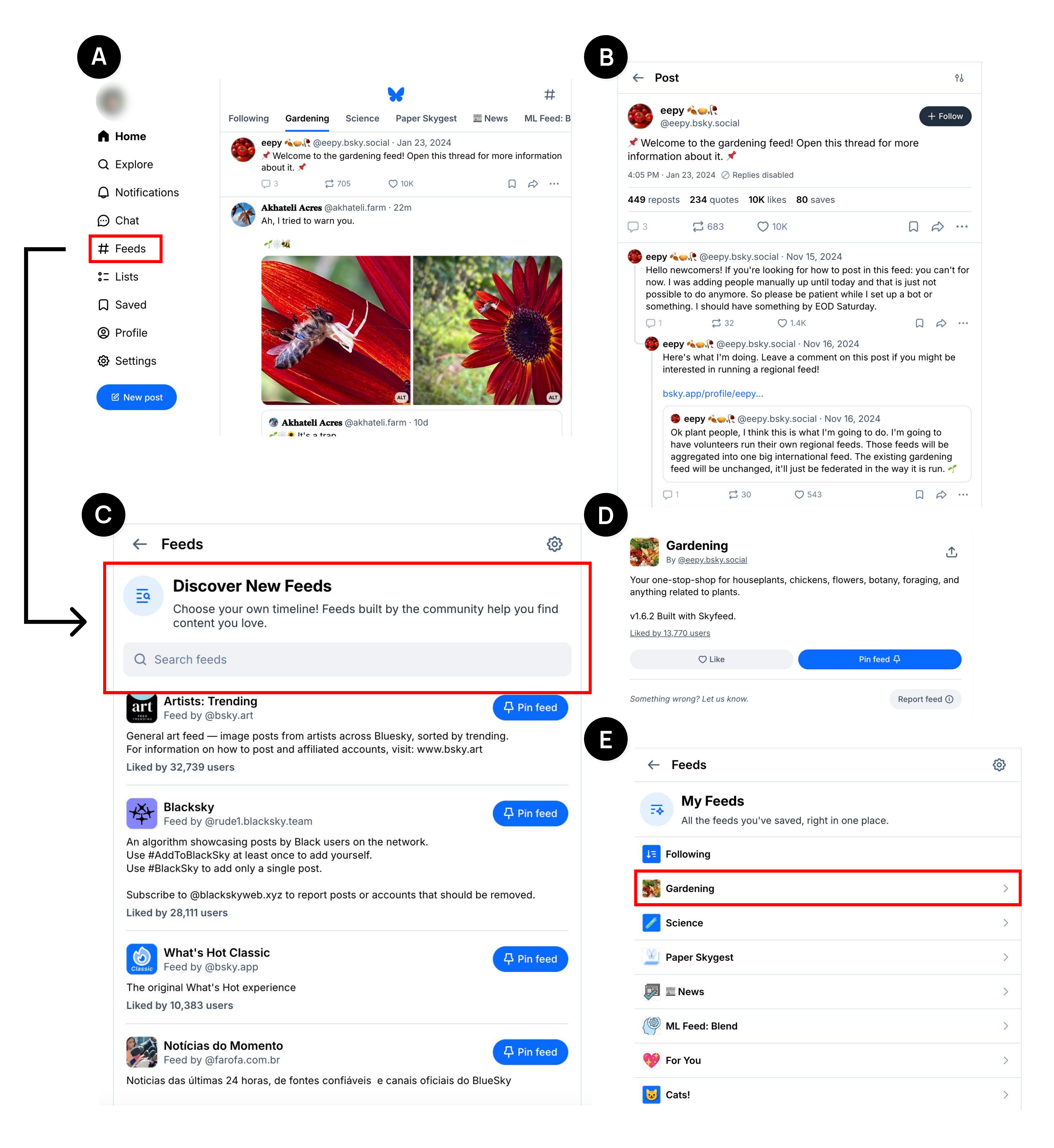}
  \caption{The \textit{Gardening} custom feed on Bluesky. Any user can (A) choose to scroll through any number of custom feeds, (B) view specific posts under the feed such as the feed creator's pinned post, (C) search for any feed to view on the platform, and (D) view the feed description, feed creator, and like, pin, or report the feed. Users can also (E) access their pinned feeds through an aggregated list from their profile. Screenshots were captured from the Bluesky web application on September 6th, 2026.}
  \label{fig:main-figure}
\end{figure*}

%May 28th, 2026

%% file: sections/02-related-work.tex
\section{Related Work}
\subsection{Algorithmic Feed Curation}
While early social media platforms deployed content in reverse chronological order~\cite{narayanan2023understanding,zhang2024form}, today, popular social media platforms such as Facebook, YouTube, and TikTok have adopted algorithmic {recommender systems} that curate content by learning from behavioral signals to maximize user engagement~\cite{boeker2022empirical, covington2016deep, backstrom2011supervised}. Despite their popularity, these systems have drawn criticism for providing users limited control over their feeds~\cite{devito2017algorithms, wang2024trustworthy, smith2022recommender}. This is exacerbated by the fact that many users are not aware of available options for personalizing their opaque feeds~\cite{hsu2020awareness, poell2020three}. Additionally, because these systems reward content that performs well by their metrics, creators are pressured to shape their posts to \textit{perform} for the algorithm rather than for their communities, a dynamic found to homogenize content across creators~\cite{lee2014social, klug2021trick} and narrow the diversity of content and communities on a platform~\cite{nechushtai2024more}.

These problems have motivated a range of alternative efforts to shift feed curation away from a platform's engagement-optimized recommender systems. 
One direction encodes human values beyond engagement into a feed's ranking objective~\cite{bernstein2023embedding, jia2024embedding}.
Because platforms have little incentive to adopt these objectives, such interventions are in practice deployed on the client side, re-ranking or rewriting the feed a platform has already delivered according to values drawn from the literature~\cite{kolluri2025alexandria} or each user's own criteria~\cite{wang2025end, rashed2025if}.
While these modifications could give users increased agency over centralized recommendations, they are fundamentally limited by the subset of content surfaced by the platform's recommenders in the first place. An adjacent line of work focuses on enabling users to build their own personal feeds from the broader pool of content available on the platform. Users can communicate their preferences through varied ways, from explicit filter controls ~\cite{bhargava2019gobo}, in-situ feedback during browsing feeds~\cite{choi2025designing}, to natural language~\cite{popowski2026social, malki2025bonsai}. Yet these interventions require both upfront and continuous configuration from users themselves as their interests evolve~\cite{barua2026compass}. Such effort is poorly aligned with the entertainment-oriented nature of feed consumption, especially since users may not have a clear articulation of their preferences.

Lastly, distributing content curation to third parties or crowdsourcing is a common but less visible way to build a feed. This ranges from curation by a single source, such as a personal blog or a followed account, to curation aggregated across many sources, as when journalists curate news feeds for their audiences~\cite{atreja2023understanding}. Crowdsourced venues such as Reddit subreddits, Stack Overflow forums, and Slashdot posts represent a further variant of this, where curation emerges from a community rather than a single curator~\cite{parnin2012crowd, chan2026examining, lampe2004slash}. More recently, these ideas have been formalized into a model for collaborative feed curation, in which users democratically delegate curation to trusted people or communities rather than to a platform whose incentives they distrust or to themselves alone~\cite{he2023cura}. These examples suggest that a promising space of intervention lies between the platform and the individual end user.

\subsection{Third Party Middleware for Social Media}
In light of these developments, scholars have long been interested in a vision of third party \textit{middleware} for social media that would offer \textit{composable} components that interoperate under an \textit{open} networking protocol~\cite{keller2021future, fukuyama2020middleware}. In theory, middleware could allow core components of social media, such as recommendation and moderation, to be hosted by third party providers and developed through standardized protocols instead of centralized platforms~\cite{ masnickprotocolsnotplatforms}. For example, a recommendation provider could access a platform's content through its open protocol and apply its own ranking logic to compile feed recommendations, while a moderation provider would return whether a post aligns with their respective moderation rules instead. 
% This allows us to choose from a variety of middleware providers, similar as to how one might choose from a variety of email clients, rather than being limited to a sole platform's imposed implementation.

The potential benefits of middleware are substantial. Middleware enables an ecosystem of competitive services where users choose between services that best fit their needs~\cite{keller2021future, hogg2024shaping}. Different communities could encode their own values into third-party services, such as a racial-justice-oriented lens from a civil rights organization, a fact-checking layer from a trusted news source, or a family-friendly version of an otherwise unfiltered feed~\cite{keller2021future}. 
This ecosystem could also result in a greater diversity of content on the platform and a proliferation of diverse communities~\cite{jhaver2023decentralizing}. 
Lastly, middleware could serve as a counterweight to misinformation and unaccountable state power, as scholars have theorized that concentrating curation and moderation in a handful of platforms has made it easier for governments to pressure those platforms into suppressing disfavored speech~\cite{fukuyama2020middleware}.

However, realizing the benefits of middleware in practice holds distinct challenges. Importantly, middleware would require lending users' data, previously centralized and secured by a single platform, to a growing number of third-party developers, introducing privacy and security vulnerabilities with these new points of access~\cite{zuckerman2024improving, masnickprotocolsnotplatforms}. Middleware is also demanding cost-wise, as third party providers would be expected to host and serve independent technical infrastructure, and it remains unclear what would incentivize them to take on this work~\cite{keller2021future}.
Perhaps most fundamentally, it is unclear whether third party middleware could compete with centralized recommenders. Today's billions of social media users are accustomed to large-scale personalized recommendation~\cite{backstrom2016serving}, and third party providers would need to build and run the algorithms and infrastructure required to match it reliably at that scale. Removing a clear centralized algorithm might therefore deteriorate the quality of users' online experiences and introduce confusion or decision fatigue~\cite{hogg2024shaping}.

% However, realizing the benefits of middleware in practice holds distinct challenges. Importantly, middleware would require lending users' data, previously centralized and secured by a single platform, to a growing number of third-party developers, introducing privacy and security vulnerabilities with these new points of access~\cite{zuckerman2024improving, masnickprotocolsnotplatforms}. 
% Middleware is also demanding cost-wise, as third party providers would be expected to host and serve independent technical infrastructure, and it remains unclear what would incentivize them to take on this work~\cite{keller2021future}.
% Even more, today's billions of social media users are accustomed to the paradigm of large-scale personalized recommendation algorithms~\cite{backstrom2016serving}.
% It is also unclear as to how third party middleware recommendation might compete with centralized recommenders. Middleware might also be demanding cost-wise for third party providers who are now expected to host and serve independent technical infrastructure, and providers' incentives to undergo this work are still yet to be determined~\cite{keller2021future}. As a result, middleware services may not be as reliable as centralized platforms that currently undertake nearly all infrastructural responsibilities. 
% Finally, removing a clear centralized algorithm might deteriorate the quality of users' online experiences on such a platform and introduce confusion or decision fatigue~\cite{hogg2024shaping}.

While the literature enumerates distinct challenges for fully realizing the ideals of middleware, today's social media platforms already exhibit roles and tools that partly align with the goals of middleware, despite not fully satisfying its tenets. 
For instance, Jhaver \etal argue that community moderators could be considered \textit{middle level} providers of governance (\eg on Reddit subreddits and Discord servers)~\cite{jhaver2023decentralizing}. Similarly, community-maintained blocklists that used to be on Twitter\footnote{To reflect the period in which these developments occurred, we will refer to the platform now known as X by its' prior name.} (now X) could be seen as providers of moderation, allowing users to filter unwanted content for others beyond what the platform moderates~\cite{geiger2016bot}. 
However, these systems are supported in-house, removing technical and cost barriers associated with self-hosting as well as many security and privacy concerns; this also leaves them more beholden to and limited by the platform.
% and marketplace extensions such as Google Chrome's Web Store enable third-party developers to act as providers of enhanced user experiences through a platform-governed API~\cite{hsu2024chrome}. 
Even fully protocol-based approaches have only partially realized the ideals of middleware: Mastodon's ActivityPub limits feed curation to personal filters and server policies rather than shared, third party, user-built infrastructure~\cite{anaobi2023will, raman2019challenges, liu2025understanding}. While some newer tools such as {Graffiti} attempt to push composable curation and moderation entirely to a client-side API to avoid infrastructure costs, they currently struggle to support complex feed recommendation algorithms~\cite{henderson2025graffiti}. More closely, Compass suggests a hybrid model in which a platform-provided base recommendation algorithm remains responsible for large-scale personalization, while users could layer lightweight continuously-updated customizations on top, pointing towards a middle ground between platform-controlled and user-created feeds~\cite{barua2026compass}.

Together, these examples show that middleware-like roles support governance and moderation in some limited ways natively on social media, yet few platforms have significantly opened up their core services to third parties. In this work, we focus on Bluesky custom feeds to study the potential of third party middleware for algorithmic feed recommendation.

\subsection{Custom Feeds and Algorithmic Choice on Bluesky}
We argue that custom feeds on Bluesky represent the first widely used instantiation of third party middleware for feed recommendation at scale~\cite{bluesky-history}. Bluesky's vision of {``algorithmic choice''} parallels the goals of middleware, aiming \textit{``to offer an open and diverse marketplace of algorithms in which communities can adapt the system to suit their needs, and users have more agency over how they spend their time and attention''}~\cite{kleppmann2024bluesky}. To do so, custom feeds are implemented on  Authenticated Transfer Protocol (\textsc{ATProto}), Bluesky's open-source decentralized networking protocol~\cite{custom-feeds-api}. We define several roles users play in supporting custom feeds within this ecosystem (see Table \ref{fig:roles}).

Particularly, custom feeds differ from prior instantiations of third-party curation (\eg journals, blogs, subreddits) in that they are not just curated by a trusted individual or community, but implemented as an explicitly shareable and publicly subscribe-able algorithm within an open protocol. Any user can create, discover, and adopt a custom feed, and any developer can build tooling which supports the functionality of custom feeds~\cite{custom-feeds-api}. Beyond other middle-level actors~\cite{jhaver2023decentralizing}, feed creators can directly author and maintain an algorithm rather than simply selecting, modifying, or creating content---making them also \textit{algorithmic providers} for communities. Bluesky also does not at the moment natively support custom feed creation within its own web application,\footnote{As of writing, we note that the Bluesky team is developing an AI-assisted feed creation tool off-platform, which is in beta testing and not accessible from the main Bluesky client~\cite{attiebluesky}.} so custom feeds must be built off-platform by third-party tooling~\cite{bluesky-algorithmic-choice}.

Due to the public nature of \textsc{ATProto}, recent large-scale quantitative analyses have confirmed that third-party tools such as \textit{SkyFeed} hosted 85\% of all Bluesky custom feeds in 2024 and that keyword-matching feeds dominate the ecosystem~\cite{quelle2025bluesky, balduf2024looking}, raising questions of what benefits custom feeds bring compared to existing similar features such as hashtags on YouTube or TikTok. Other work has investigated how users leverage general Bluesky moderation practices, such as starter packs for bootstrapping social media migration, and moderation practices for individual users and communities through vetted blocklists~\cite{sokoto2026open, balduf2025bootstrapping, bono2026self, nogara2026longitudinal}. 
However, prior work has not built a concrete understanding of feed creators themselves. We understand little about who are feed creators, how they build and maintain custom feeds, and why and how they sustain their custom feeds over time. Feed creators are key to Bluesky's vision of algorithmic choice~\cite{bluesky-algorithmic-choice}, yet they remain the least understood role in the custom feed ecosystem; thus, our work aims to fill this gap.

\input{figures/roles}

%% file: figures/roles.tex
\begin{table}[]
  \centering
  \small
  \setlength{\tabcolsep}{4pt}
  \renewcommand{\arraystretch}{1.3}
  \caption{Definitions of four key roles within Bluesky's custom feed ecosystem: feed creators, feed contributors, feed consumers, and third-party developers. We order these roles top-down by their level of involvement in an individual custom feed.}
  \begin{tabular}{p{3cm} p{10cm}}
  \toprule
  \textbf{User} & \textbf{Role} \\
  \midrule
  \textbf{Feed Creator} & A user who builds, maintains, and sustains a custom feed, either through using a third-party tool, or directly via the \textsc{ATProto} API. \\
  \textbf{Feed Contributor} & A user whose content appears in a custom feed, either {knowingly} (\eg by including a designated hashtag in their post) or {unknowingly} (\eg when their posts match a keyword filter set by a feed creator). \\
  \textbf{Feed Consumer} & A user who pins, likes, or otherwise views a custom feed's content without necessarily contributing to it. Akin to a subscriber. \\
  \textbf{Third Party \newline Developer} & A developer who builds and maintains third-party tools supporting custom feed creation (\eg \textit{SkyFeed}, \textit{Graze}, \textit{Bluesky Feed Creator (BFC)}). Distinctly, these are not feed creators who implement their own infrastructure for custom feeds, but third party feed building service providers across the ecosystem. \\
  \bottomrule
  \end{tabular}
  \label{fig:roles}
\end{table}

%% file: sections/03-method.tex
\section{Method}

\input{figures/study-specifics}
\input{figures/third-party-devs}

We conduct a mixed-methods study, with qualitative analysis of semi-structured interviews with feed creators and third party developers of feed tools, coupled with a large-scale quantitative analysis of all publicly accessible custom feeds on Bluesky.

\subsection{Feed Creator Interviews}

We conducted interviews over Zoom with $n = 24$ Bluesky feed creators and $n = 2$ third party developers. All participants were compensated with an online gift card at a rate of \$20 USD per 45 minutes and \$30 USD per 60 minutes, except four participants who voluntarily declined compensation. Our study protocol was approved and deemed exempt by our institutional review board.

\subsubsection{Recruitment}

We recruited participants primarily on Bluesky itself through two calls for participation posted in April 2025 and October 2025, which spread naturally through dissemination on the platform. We also posted calls for participation on Discord servers and Reddit subreddits that facilitate discussion related to Bluesky. Based on a public list of top feeds,\footnote{https://blueskydirectory.com/feeds/all} we also directly messaged or emailed feed creators.
% Our call for participation included a sign-up survey that collected basic demographic information and consent to participate in our study. 
We pre-screened participants to ensure that all participants held experience in creating custom feeds as a feed creator or developing third party tools for custom feeds. 

We designate each participant a numerical ID according to their role of feed creator (C1--C24) or third party developer (D1--D2).
Our set of feed creator participants maintained a diverse range of custom feeds. Three had 10,000+ likes, seven had 1000+ likes, nine had 100+ likes, and five had $<$ 100 likes. Feed topics varied from casual hobbies (\eg art, gaming) and fandoms (\eg anime, streaming) to professional communities (\eg science, news reporting). Our two third party developer participants built two of the most widely used third party tools on the platform; specifically, D1 is the lead backend developer of one and D2 is the solo full-stack developer of the other.
We report participant details in Table~\ref{study-specifics} and demographics in Appendix ~\ref{appendix:participants}.
\subsubsection{Study procedure}

Each interview was conducted by two study team members, a primary interviewer and a notetaker, with one team member serving as primary interviewer for the majority of interviews. In each interview, we discussed participants' motivations for joining Bluesky, their experiences creating and growing their feeds, and thoughts they had on the future of custom feeds and Bluesky as a whole. We detail our semi-structured interview protocol in Appendix~\ref{appendix:interviews}.

After all interview sessions were conducted, one study team member downloaded each meeting transcript as a raw text file from Zoom, lightly edited each transcript for clarity, and thematically coded participant responses~\cite{braun2006using} on an online whiteboarding tool.\footnote{https://figma.com/figjam} Three study team members met weekly to discuss emerging themes and align on progress. 

\subsection{Data Collection}\label{section:data-collection}

In alignment with related work~\cite{quelle2025bluesky, balduf2025bootstrapping}, we built our custom feed dataset by scraping all accessible \textsc{ATProto} user decentralized identifiers (DIDs) from 146 known personal data server (PDS) instances\footnote{https://github.com/jazware/bsky-experiments} through the Bluesky Relay in January 2026. We used the \texttt{atproto.sync.listRepos}\footnote{https://docs.bsky.app/docs/api/com-atproto-sync-list-repos} endpoint, yielding a set of $33,199,368$ unique DIDs representing 31.8M active and 1.4M inactive accounts. While public statistics at the time tracked $\sim$40M+ total Bluesky users, this discrepancy may stem from accounts hosted on PDS instances outside our set of 146, or accounts registered but not yet propagated through the Relay at our time of collection.
We then directly scanned \textsc{ATProto} repositories and extracted custom feed records created by users from our set of DIDs using the \texttt{getActorFeeds}\footnote{https://docs.bsky.app/docs/api/app-bsky-feed-get-actor-feeds} endpoint. We collected the feed DID, the feed creator's DID, the feed creator's domain handle, the feed creation date, and likes at the time of collection. These requests ran asynchronously on four NVIDIA 2080ti GPUs via our university cloud computing cluster, ultimately yielding $n=97,719$ unique feed DIDs. We provide a sample JSON object of each feed's metadata we collected in Appendix ~\ref{appendix:json}.

\subsubsection{Collecting custom feed metadata}\label{collecting_metadata}

To collect further metadata from our set of custom feed DIDs, we performed a second large-scale data collection using publicly accessible endpoints from \textsc{ATProto} API. For each of our $n=97,719$ feed DIDs, we collected (i) the complete set of users who had liked the feed, (ii) metadata for up to the 500 most recent posts in each feed, including authorship, timestamps, replies, quotes, media, languages, and engagement statistics, and (iii) the set of users who interacted with those posts through likes and replies. We use users who liked a feed as a proxy for feed subscribers. This is our best proxy of a traditional \textit{subscriber} because  \textsc{ATProto} currently does not provide any signal as to whether a user has pinned or viewed a feed explicitly.

For each feed, we then identified the unique sets of users who had either liked or replied to the post as \textit{post likers} and \textit{post repliers}. We define an \textit{active consumer} as a user who appears in both i) the feed's subscriber base and ii) the set of all unique users who have liked a post on the feed within the 500-post window. We additionally formulate an \textit{active consumer participation rate} for each feed as the intersection of post likers and post repliers over the total feed likers:

\[\text{active consumer participation} = \frac{\text{Feed Likers} \cap (\text{Post Likers} \cup \text{Post Repliers})}{\text{Feed Likers}}
\]

\subsubsection{Custom feed dataset and summary statistics}

After discarding feeds that did not return any metadata, our final dataset comprised of $n = 88,302$ custom feeds created by $n = 41,955$ unique feed creators. The median creator maintains one feed (with mean being $2.1$ feeds per creator). Of our feeds, $49,069$ had at least one subscriber at our time of collection, with $1,319,921$ total likers for all feeds. The median feed had just $3$ subscribers, while the mean was $26.9$ and the maximum was $50,757$. The majority of feeds ($77.0\%$) had between $1$ and $10$ subscribers, $20.2\%$ had between $11$ and $100$, $2.5\%$ had between $101$ and $1,000$, and only $0.2\%$ ($n = 121$) exceeded $1,000$ subscribers (see Appendix~\ref{appendix:descriptive_statistics}). These summary statistics are consistent with prior quantitative analyses that suggest the custom feed ecosystem largely consists of smaller individual feeds with a handful of high-visibility feeds~\cite{quelle2025bluesky}. This dataset enabled us to characterize creator activity, consumer participation, and the structural properties of custom feeds at ecosystem scale paired with our qualitative findings.

%% file: figures/study-specifics.tex
\setlength{\tabcolsep}{4pt}
\renewcommand{\arraystretch}{1.2}
\begin{table*}[!htbp]
  \centering
  \scriptsize
  \caption{We lay out our 24 feed creator interview participants, including the third party tool they used, intended feed purpose, orientation, feed likes, feed logic stated in interviews, and a description of their feed. Creators who implemented their own custom feed infrastructure and/or feed logic instead of using a common third party tool are marked as N/A.}
  \begin{tabular}{l p{1cm} p{1.3cm} p{1cm} p{1.3cm} p{6cm}}
  \toprule
  \textbf{ID} & \textbf{Tool} & \textbf{Orientation} & \textbf{Likes} & \textbf{Feed Logic} & \textbf{Feed Description} \\
  \midrule
  \textbf{C1}  & N/A          & Community     & 1000+  & Keywords & A custom-built feed serving scientists and science communicators, with a team of user moderators for specific subfields of expertise. \\
  % \textbf{P2}  & N/A                  & N/A       & N/A    & A developer who built and maintains a no-code, block-based third-party feed hosting platform supporting hundreds of feeds on Bluesky. \\
  %with emerging monetization features for feed creators.
  \textbf{C2}  & N/A        & Utility     & 100+   & Personalized \newline algorithm  & A custom-built personalized recommendation feed that draws on user-specific data by custom implementation. \\
  \textbf{C3}  & \textit{SkyFeed}              & Community & 100+ & Keywords  & A fandom feed using hashtag-based keyword matching, actively curating fanart and community content. \\
  \textbf{C4}  & \textit{Graze}                & Community & $<$100 & Keywords & A fandom feed for a popular video game, migrated from \textit{SkyFeed} to \textit{Graze} to support community features. \\
  \textbf{C5}  & \textit{Graze}                & Community      & 1000+  & Keywords & A hobbyist nature topic feed using emoji and hashtag keyword matching, managing spam and self-promotion. \\
  \textbf{C6}  & \textit{SkyFeed}              & Utility     & $<$100 & Keywords & A simple political topic feed to catch discussion of these posts, primarily set-it-and-forget-it in their maintenance approach. \\
  \textbf{C7}  & \textit{BFC} & Community      & 1000+  & Keywords & An cross-disciplinary feed of multiple topics welcoming community contributions and enforcing content standards. \\
  \textbf{C8}  & N/A          & Community & 10000+ & Keywords, User Lists & A custom-built community feed using a self-hosted algorithm and infrastructure, enabling users to self-add and self-post via hashtag. \\
  % \textbf{P10} & N/A                  & N/A       & N/A    & A developer who built and maintains a widely-used keyword-matching third-party feed creation tool, supporting hundreds of feeds on Bluesky \\
  \textbf{C9} & \textit{SkyFeed}              & Utility     & 10000+ & Keywords, User Lists & A high-traffic curated news feed, focusing on delivering quality journalism and curated content.  \\
  \textbf{C10} & \textit{SkyFeed}              & Community     & 100+ & Keywords  & A scientific-adjacent feed drawing from contributor-curated images and publications. \\
  \textbf{C11} & \textit{SkyFeed}              & Community & 1000+ & Keywords  & A literary fandom community feed, attracting popular power users and fostering fandom discussion. \\
  \textbf{C12} & \textit{Graze}                & Community & $<$100 & Keywords & A streamer fandom feed, displayed live on stream and requires DM-based onboarding in a bilingual community. \\
  \textbf{C13} & \textit{SkyFeed}              & Community      & 100+ & Keywords  & A board game and fighting game community, actively welcoming newcomers who discover the feed organically. \\
  \textbf{C14} & \textit{Graze}                & Community & 1000+ & Keywords, User Lists & A popular crowdsourced social feed where users trigger inclusion and discussion by replying to posts with a specific keyword. \\
  \textbf{C15} & \textit{Graze}                & Utility     & 10000+ & Keywords & A high-traffic news feed based on a vetted allowlist of verified organizations, monetized transparently with ads. \\
  \textbf{C16} & \textit{Graze}                & Community      & 1000+ & Keywords & An art feed which surfaces content from new creators under a follower count, enforcing a no-AI policy. \\
  \textbf{C17} & \textit{BFC} & Community      & 100+ & Keywords  & A gaming feed with active livestreaming events, onboarding new members and using labeler functionality to manage AI-generated content. \\
  \textbf{C18} & \textit{BFC} & Community     & 1000+ & Keywords & A hobbyist image-forward feed using emoji-based keyword matching, with a set-it-and-forget-it maintenance approach. \\
  \textbf{C19} & \textit{Graze}                & Community      & 100+ & Keywords  & A professional academic feed using custom labeler functionality, serving multilingual and specialized audiences. \\
  \textbf{C20} & \textit{Graze}                & Community      & 100+ & Keywords  & An art feed using hashtags, enforcing strict content standards including a no-AI/NFT policy with banning for violations. \\
  \textbf{C21} & \textit{Graze}                & Community      & 100+ & Keywords  & A  children's illustration feed using hashtag keyword matching, primarily used by professional illustrators and publishers. \\
  \textbf{C22} & \textit{SkyFeed}              & Community     & $<$100 & Keywords & A scientific feed using hashtag keyword matching, which has seen limited growth since its' creation. \\
  \textbf{C23} & \textit{BFC} & Community     & $<$100 & Keywords & An art subgenre feed for photography and artwork with minimal ongoing maintenance and contributors posting sporadically. \\
  \textbf{C24} & \textit{Graze}                & Community      & 100+ & Keywords  & A gaming fandom feed covering fanart and community events; creator actively onboards and promotes the feed to new community members. \\
  \bottomrule
  \end{tabular}
  \label{study-specifics}
\end{table*}

%% file: figures/third-party-devs.tex
\setlength{\tabcolsep}{4pt}
\begin{table*}[!htbp]
  \centering
  \scriptsize
  \caption{We lay out our two third party developer participants, who had built two of the most widely used third party tools on Bluesky.}
    \begin{tabular}{l p{2cm} p{4cm} p{6cm}}
    \toprule
    \textbf{ID} & \textbf{Feeds Hosted} & \textbf{Tool Features} & \textbf{Tool Description} \\
    \midrule
    \textbf{D1} & 1000+ & block-based algorithm building, moderation support, monetization, ML rankings & A free third-party tool supporting keyword matching and machine learning ranking, with monetization options for feed creators. \\
    \textbf{D2} & 1000+ & keywords, regular expression-based algorithm building, labeler support & A free no-code, regular expression-based feed creation third party tool responsible for a sizable percentage of all custom feeds on Bluesky. \\
    \bottomrule
  \end{tabular}
  \label{third-party-devs}
\end{table*}

%% file: sections/04-findings.tex
\section{Findings}

\subsection{Characterizing Feed Creators and Feeds}

\input{figures/findings_figure}

\subsubsection{\textbf{Two ways feed creators see their role: utility-providing and community-building}}\label{ref:4.1}

Through our interviews, we inductively derived two orientations that custom feed creators expressed towards their custom feeds. We term these \textit{utility-providing} and \textit{community-building}. Each orientation captures how creators understand their role, how they curate their feeds over time, and what their feeds' success means to them. Below we define and characterize these orientations in more detail.
% We use these orientations as an organizational lens to contrast creators' experiences with custom feeds.
\begin{itemize}
\item
\textbf{{\textit{Utility-providing}} creators treat their feed primarily as a content delivery mechanism for consumers.} In their feeds, the typical consumer is foremost a \textit{follower} of the feed and need not \textit{contribute} posts to the feed's content. A \textit{utility-providing} feed creator desires to surface relevant content on Bluesky to the feed, and feed contributors may not even be aware that their content is being sourced for the feed. 
% Furthermore, we identify that in feeds created with this orientation in mind, consumers engage more with the content within the feed itself rather than interacting among other consumers. 

\item
\textbf{In contrast, \textit{community-building} creators support feeds that foster consumers interacting with each other.} In these feeds, feed \textit{consumers} themselves are the primary \textit{contributors} to the feed. Instead of serving simply as a source of information, these feeds support communities around shared interests. In many cases, they support existing communities that predated Bluesky, helping consumers find each other on a new platform.
\end{itemize}

We note that these orientations loosely map onto prior theoretical work on community types, such as \textit{bond-based} versus \textit{interest-based} communities~\cite{ren2007applying}. At the content level, utility feeds function similarly to traditional \textit{communities of interest}~\cite{fischer2001communities} or \textit{networked publics}~\cite{boyd2010social}, optimizing for information routing, discoverability, and passive consumption across a large audience. In contrast, community feeds operate closer to \textit{communities of practice}~\cite{wenger1999communities} or \textit{intimate publics}~\cite{berlant1998intimacy}, relying on mutual engagement and a shared social repertoire. Furthermore, this distinction extends to the feed creators themselves. Drawing from Seering \textit{et al.}'s metaphors in moderation~\cite{seering2022metaphors}, community-building creators seem to embody values of a \textit{Gardener} as tending to a community, while utility-providing creators function closer to the \textit{Filter}, as engaging less with community members and oriented towards screening out problematic content or users.

% More closely, Seering \etal delineate metaphors in moderation~\cite{seering2022metaphors}, where community-building feed creators embody a \textit{Gardener} where conversely, utility-providing creators function closer to automated infrastructure.

% We note that these orientations loosely map onto prior theoretical work on community types, such as bond-based versus interest-based communities~\cite{ren2007applying}. While \textit{utility-providing} feeds aggregate content around a shared {interest or topic}, \textit{community-building} feeds can successfully foster {bonds} among consumers with shared interests. One difference is that we use the distinction  to focus on characterizing feed creators and their orientation toward their role as opposed to communities.

\subsubsection{\textbf{Feed creator tasks and measures of success: ``set-it-and-forget-it'' versus community steward}}

Unsurprisingly, the two feed creator orientations have distinct ideas about the tasks they should take on. \textit{Utility-providing} creators passively view their feed as observers or focus their efforts on curating content, without asking subscribers for feedback. Creators described this approach as \textit{``set it and forget it''}, where they configure the feed's logic upfront and then step back. As a result, these creators measure success largely by whether the feed does its job of surfacing relevant high-quality content. 
Some \textit{utility-providing} creators additionally described seeking to improve information access for their subscribers. For example, C9 maintains a news sourcing feed that provides access to articles behind paywalled publishers. Their role \textit{``is not to make money or get personal attention''} but instead to \textit{``enable people to connect with actual news, as opposed to being captured by whatever happens to be free and pushed to them.''}

\textit{Community-building} creators instead act as the steward of spaces where consumers can interact \textit{with each other} through the feed. For example, C14 maintains a feed where consumers directly use a hashtag to forward content to the feed; C3's artwork feed welcomes contributions from consumers themselves to share their artwork with each other, not just sourcing posts via hashtag filters from across the network. This role require more sustained involvement, including tasks such as onboarding new members, moderating content, and actively participating in the feed themselves:

\begin{quote}
    \textit{``I almost see myself as the librarian. If we're in a library, I'm the librarian behind the desk. I'm making sure people aren't yelling, they're picking up their trash, and putting books back where they're supposed to be. I'm making the place as presentable and welcoming as possible.''} \hfill --C19
\end{quote}

\textit{Community-building} creators measure success relationally, \eg whether consumers found each other, felt welcomed, or discovered a community they didn't know existed on the platform. \textit{``I feel achievement in the fact that it gives people looking to find the [feed topic] community a pretty quick and easy way to do it, without having to be included in a starter pack or a list. It's open for basically anybody to be like, "Oh, I like [feed topic]," and get their post seen. That part, I do feel proud of''} (C11).

\subsubsection{\textbf{How feeds got started and evolved: from replication and migration from other platforms to new experiments in feed curation}} We found that many feed creators got their start in one of two ways.
First, in the early days of Bluesky, creators explained that it was difficult to find content that catered to their specific interests. As such, they made feeds purely as an easier way to curate content for themselves, without initially seeking to share with other users: \textit{``I was just excited by the idea of having control, so I would create, unpublish, and delete feeds just for the hype of it.''} (C6).
They then realized that because their feeds are public, they could curate content not just for themselves but also for other users. 
Coupled with Bluesky moving out of invite-only access, creators recognized that they could use feeds to replicate features from existing social media platforms (\eg Twitter, Reddit, Discord). 
Soon, these creators began expanding their feeds towards wider audiences.
Many feeds that got their start this way have a feed logic that relies on simple keyword matching, where any post containing a chosen term or hashtag is automatically surfaced (\eg C5, C18).
Over time, other feeds have adopted a more deliberate curation: feeds that filter posts only from a vetted list of contributors (C1), surface high-quality journalism from selected sources (C9), or even personalize content for individual consumers (C2). Together, these forms span a spectrum from feeds that broadcast a creator's personal interests platform-wide to feeds whose value comes from careful or complex curatorial gatekeeping.

% (C3, C4, C7, C8, C11, C12, C13, C14, C16, C17, C19, C20, C21, C24)
While the first way maps onto utility-providing feed creators, the second way maps to creators seeking to build or recreate community. 
As Bluesky is a relatively new platform, many creators initially sought to migrate existing communities from other platforms onto Bluesky. 
For example, C11, who maintains a reading community, stated: \textit{``The community existed long before social media---back in forums, even before the internet was widespread. But specifically, I was trying to court the people from [genre] Twitter and help them find each other again, and also just to capture anybody who likes [genre] novels.''}
Over time, as the community grew, creators also sought to bring in new members.
For instance, C13's feed catches posts with any keywords related to their hobbyist board game community. Since this catches any discussion on Bluesky related to their topic, the feed can draw engagement from new contributors:

\begin{quote}
    \textit{``The great thing about these feeds is that even if you don't know about the feed, your posts are still in it. So somebody will be like, "I want to learn about [board game]," and anyone can just jump in, and this complete stranger has found us and is invited into the community.''} \hfill --C13
\end{quote}

Some \textit{community-building} creators devised new social mechanics to shape how people participate in their feeds. For example, C17 built a feed that surfaces content from contributors below a certain follower threshold, where a community of users then amplifies content from these new users rather than established ones on Bluesky. C14 took a different approach, building crowdsourced feeds where any consumer could take part in feed curation: if you replied to a post with a specific keyword, that post would automatically go to the feed,  leveraging comment engagement itself as a curation mechanism. 
% These mechanics represent a fundamentally different relationship between how consumers interact in their feeds to form community in these feeds in contrast to \textit{utility-providing} feeds.

% \subsubsection{\textbf{Custom feed content extends beyond both feed subscribers and consumers}}

% We further sought to understand where feed content originates and how it reaches audiences. To do so, we analyzed feed authorship distribution and the sources of post engagement across our set of custom feeds (see Figure ~\ref{contributions}).

%NOTE TO REMOVE
% \input{figures/contributions}

\input{figures/active_consumers}

\subsubsection{\textbf{Feed creator orientations as a lens onto differing user activity around feeds}}

% After qualitatively defining the \textit{utility-providing} and \textit{community-building} archetypes of custom feeds in our interviews, we sought to formalize this taxonomy at scale in the custom feed ecosystem. To do so, we analyzed our custom feed dataset \S \ref{section:data-collection}.

\input{figures/interactions_engagement}

The two main feed creator orientations provide a useful lens for us to better understand differences in activity around feeds.
We find that many custom feeds on Bluesky operate without active participation from feed creators or consumers. Across $n = 88,302$ feeds, 67.5\% ($n = 59,582$) did not include any posts authored by the feed creator in the last 500-post window (see Figure ~\ref{active_consumers}). 
% This suggests that while many creators might not include their own content in their own feed algorithms, there still exists a subset of feeds which do.
We also found that most feeds had no active consumers (78.8\%), by our definition in \S\ref{collecting_metadata}.
% In the proportion of feeds that did have active consumers, we find that a small proportion of feeds have active consumers and partially active consumers, where the majority do not. 
While not always the case, this behavior more closely aligns with the \textit{utility-providing} orientation, where the creator produces the feed algorithm but does not create content for the feed or seek to foster engagement among consumers. This suggests that most feed creators are operating from a goal of providing utility.

We also find that feeds with an active creator see higher levels of feed interaction. We define a feed's \textit{average interaction per post} as the mean number of total interactions (\eg likes, reposts, and replies) across up to 500 of the feed's most recent posts. Feeds with an active creator generated substantially higher median average interactions per post ($\text{median} = 11.94$, $IQR = [4.00, 36.32]$) compared to those with inactive creators ($\text{median} = 1.56$, $IQR = [0.00, 9.22]$) (see Figure ~\ref{interactions_engagement}). We confirm via a Mann-Whitney U test that this difference in engagement is statistically significant ($p < 0.001$). 
% Notably, the maximums for both groups were driven by feeds with very high levels of engagement, some of which reached ~13,000 average interactions for active creators and ~130,000 for inactive creators. 
This suggests that higher levels of engagement occur in feeds where the feed creator is actively participating (\ie where the creator likely sees themselves as \textit{community-building}).

\subsection{Building Custom Feeds}
While creators differ in their orientations and motivations for building their feeds, we found that they largely rely on the same set of third-party tools to build them, with technical complexity scaling up as curation needs grow. Table~\ref{tab:thirdpartytools} summarizes these tools and their features.

\input{figures/third-party-tools}

\subsubsection{\textbf{Nearly all feeds use keyword matching through third party tools}}

The majority of feed creators we interviewed ($n = 21$) used a third party tool to create their feed or had eventually migrated onto a third party tool. Aligning with work in 2024 that found \textit{SkyFeed} hosted 85\% of all feeds on the network~\cite{quelle2025bluesky}, creators explain that third party tools are simply the easiest and best way to create custom feeds on Bluesky, citing their favorites as \textit{SkyFeed}, \textit{Graze}, and \textit{BFC}~\cite{skyfeed, graze, blueskyfeedcreator}. Creators describe that these tools offer customizable ways to add functionality to their feeds, and some creators chose specific third party tools for their intended use cases. C17, who built a community gaming feed, determined:
\textit{``As much as I loved it, \textit{SkyFeed} didn't have the security I wanted, especially for a growing community. I didn't just want a place for people to post; I wanted a place where I could curate certain things that shouldn't be in the feed and block users. So I decided to use \textit{BFC}. With that, you have terms, you have labels, and you can completely block bad actors from interacting with the feed. You can also link all your feeds, so if you block someone in one, you can block them in all.''}
This flexibility stems from the \textsc{ATProto}'s dynamic between third party developers, creators, and Bluesky as a platform itself.
Creators have the freedom to choose from a range of options of interoperating third-party middleware providers, which differ from prior approaches: 
\textit{``It would be cool if more things were built into Bluesky, but I like that it's flexible. Having that control is worth the trade-off of less direct support.''} (C16)   

While the actual feed logic for many custom feeds is commonly through defining keyword matching regular expressions (regex), creators evolved beyond this to designate lists of users whose posts would appear on the feed, as well as labeler-based features. Nontechnical feed creators also enumerated challenges with learning to use regex. Some creators fully admitted that while they had derived a basic working regex filter, they still were unsure of how the expressions exactly worked and therefore aim to touch their feed settings as little as possible (C21). 

\subsubsection{\textbf{Creators devised functionality beyond keyword matching to enhance custom feeds}}

For creators who needed more precise control over who could post in their feed, keyword matching alone proved insufficient. Some creators evolved to maintaining vetted lists of approved accounts as their primary feed logic, such as a collective workout accountability group where the feed creator actively adds contributors by request (C18), or a news feed restricted to verifiable journalism organizations (C15). {Unlike keyword matching, which surfaces any post matching a string, list-based feeds give creators control over \textit{whose} contributions appear.}

As feed creators grew familiar with third party tools, some developed more specific content filtering mechanisms to complement their feed building processes, such as for AI-generated imagery and misinformation. For example, C17, a feed creator who uses \textit{Graze} for their feeds, built a popular labeler for AI-generated imagery, where their aim is \textit{``to always be neutral and factual: either it is or is not AI''} (C17).

Many feed creators also desired more features from third party tools. While third party developers do try to meet all feed creators' needs, some tasks can become unwarranted responsibilities for them. Specifically, D2 stated that they had to implement features that they originally did not want to but had to for the community, such as caching posts beyond seven days. D2 states, \textit{``I hate this feature because it makes my life harder. I have to worry about backups. But the demand was overwhelming.''} 

Some creators who did not have technical expertise but wanted to improve their feed logic beyond keyword matching had to ask for help to devise custom functionality to support their feeds. For example, while C19 was satisfied with their current level of feed functionality, their community required specific labeling functionality to which they had to reach out for help with: \textit{``The biggest limiting factor was for the labeler. There's no simple tool for that like there is for feeds. It requires coding, so I had to ask people for assistance. But I'm comfortable with the technical expertise I have to run things currently. The community is looking for a basic place to talk and network, and to limit misinformation and harassment.''} 

\subsubsection{\textbf{Some creators find value in implementing their own feed infrastructure}}

While third party tools are used ubiquitously, three feed creators (C1, C2, C8) explained that they built their own custom infrastructure to support their custom feed. This was because third-party tools could not meet their needs; these creators wanted to operate on larger slices of available \textsc{ATProto} data and build more sophisticated logic for their custom feeds beyond the capabilities of existing third party tools.
For example, C2's feed is a personalized feed that draws from signals on who a feed consumer follows. They precompute personalized recommendations for each individual user's feed and transitioned their storage from a serverless AWS Lambda instance to a hosted AWS EC2 cloud server to keep costs low. The freedom of working without a third-party tool and building directly with the \textsc{ATProto} provides C2 the ability to do \textit{``real recommendation work''}:

\begin{quote}
    \textit{``Recently, they made a choice to release interaction data to custom feed developers. Previously, we only knew that a user "liked" a post and that they used our feed around that timestamp, but we didn't know for sure if they liked it while on our feed. Now, Bluesky sends us the same fine-grained data they use for their \textsc{Discover} feed. This is a big difference because it allows us to do real recommendation work.''} \hfill --C2
\end{quote}

C8 faced a different limitation, as their feed logic was too complex to express through block-based tooling, requiring custom ranking, video support, trending detection, and community membership logic that no existing tool could handle together. They built their feed from scratch in Rust, implementing their own open-source \textsc{ATProto} client in the process, stating that they \textit{``had to scale the infrastructure to handle a massive increase in traffic, especially during major events, when the feed had to serve a million people at once''} (C8). 

These feeds illustrate that third party tools do not work for everyone. Creators always have the ability to implement logic beyond the relatively simple functionality that third party tools provide. Feeds that began as community projects like C8's have grown into independent infrastructure, complete with moderation teams and protocol-level implementations that can function independently of Bluesky itself.

\subsection{Maintaining Custom Feeds}

Creators describe custom feed maintenance as ongoing work that introduces tasks specific to the feed creator role, extending beyond what prior research on online community moderation has documented~\cite{seering2022metaphors}.
% These tasks include setting and enforcing rules, onboarding new members, and responding to community feedback. 
Specifically, we identified two kinds of maintenance tasks: \textit{proactive} tasks (\eg evaluating analytics, updating feed logic) and \textit{reactive} tasks (\eg moderation, responding to user feedback). Notably, for \textit{utility-providing} creators, their maintenance is infrequent, consistent with the common set-it-and-forget-it philosophy (\S \ref{ref:4.1}). Community-building creators, by contrast, engage in substantially more ongoing maintenance work across both categories. 

\subsubsection{\textbf{Feed moderation is supported by labelers and often anticipated beforehand.}}

In \textit{utility-providing} feeds, a feed creator's moderation is largely reactive. Because \textit{utility-providing} feeds focus on curating content from contributors that the feed creator is already familiar with, unwanted posts and rule violations can be anticipated beforehand and dealt with by exclusion criteria. Creators add block lists for known bad actors and connect feeds to third-party labelers that pre-screen content by type such as spam, unwanted ads, AI-generated imagery, or misinformation. 
Before these moderation features existed, creators oftentimes had to \textit{``manually find keywords and add hundreds of accounts to a blocklist to filter them out.''} (C3), but now many creators have adopted these strategies across their feeds. As C24 explained, \textit{``If someone makes a new account to spam, it takes me one second to add them to the filter list. There are people who use the top 10 hashtags on every post to sell stuff like Amazon links. I filter those out by putting popular hashtags in the removal list if they appear alongside spam keywords.''} \hfill --C24

In contrast, \textit{community-building} feeds require more nuanced moderation. Because these feeds thrive on user interaction, moderation decisions carry social weight that content-based exclusions do not. Creators like C12 describe preferring corrective actions over permanent bans, recognizing that their relationship with community members matters more beyond a single violation: \textit{``On Twitter, removing someone from the community was a permanent, heavy-handed action. On Bluesky, I can remove a contributor from my feed's list if they post something inappropriate, which removes their posts, but I can also add them back. This allows for a discussion, which is a big improvement.''}

\subsubsection{\textbf{Community-building creators put in more active effort to maintain their feed community}}~\label{creator-effort}
% While many simple \textit{utility-providing} feeds tend to not have direct daily maintenance tasks to tackle, creators of \textit{community-building} feeds must put in active effort to maintain their feeds. 
Some community-building creators describe proactive promotion and engagement as an essential task to maintain a healthy feed, rather than waiting for potential consumers to organically discover their feed. 
% For example, C17, who maintains a growing streaming community, states that they can \textit{``recognize 3,000 accounts alone''} who have visited their feed. 
While browsing other feeds on Bluesky, they actively scan their custom feed to detect new users and continue \textit{``onboarding new people to this day''} by reaching out and introducing themselves as the feed creator. More extremely, C24 directly markets their fandom feed on Bluesky to potential consumers by persistently engaging with any relevant content on Bluesky:
\textit{``I have been commenting on every single [fandom] post I find, telling them about feeds and how to set up their Bluesky accounts. I've sent thousands of messages. It works; people tell me, "I stayed on Bluesky because of your post." But obviously, that is not a sustainable strategy.''}

%C12, C13, C14, C22, C23, C24)
Creators voluntarily choose to do this because they have found the \textit{feed discovery process} ineffective and largely unsupported by Bluesky. C24 and six other creators expressed that Bluesky should support {feed discovery} as \textit{``new users on Bluesky have no idea that you can create or subscribe to custom feeds''} (C24). For new users, other features on Bluesky (\eg the \textsc{Discover} feed, user lists, and starter packs) compete for their attention, and that Bluesky as a platform fails to communicate the potential of custom feeds as an alternative to these more traditional social media features (C12, C20). Creators like C12 criticized Bluesky's feed discovery as stagnant:
\begin{quote}
\textit{``The feed discovery inside Bluesky itself is not great. It's stuck in time. If you create a new account and go to the ``Feeds'' tab, you'll see the same top feeds from two years ago. They are all in English and often about the U.S., so it's not interesting for people outside that. This is one of my biggest pains.''} \hfill --C12
\end{quote}

% Creators believed this stems from other features on Bluesky competing for consumers' attention (C12, C20). Creators described that features such as the \textit{Discover} feed and starter packs held distinct use cases that were more separate than custom feeds: \textit{``The feed felt more about community, and starter packs became more about status. A starter pack is just a collection of accounts that you think people should be interested in. I had been on for a year or two, so I knew most of the people who had come over, and I was just trying to quickly put everybody in a starter pack for people to pass around. But then it started to feel more like a status thing and less like a "find community" thing.''} \hfill --C11

\subsubsection{\textbf{{Feed creators are responsive to feedback but rarely receive explicit feedback.}}}

All creators state that feedback of any kind is largely uncommon. To maintain their feeds, they must anticipate both \textit{implicit} and \textit{explicit} feedback from {contributors} and {consumers} alike. 

\begin{itemize}
    \item \textbf{\textit{Understanding implicit feedback}}. Because direct user feedback is rare, creators use third-party analytics as an implicit signal of feed health. This includes feed post likes, daily active users, total volume of posts captured by the feed (\textit{Graze}), how often a feed is loaded (\textit{SkyFeed}), and how many times the feed is opened (\textit{BFC}). While this helps to get a sense of growth over time, creators recognize that this isn't a directly actionable way to maintain their feeds: \textit{``These metrics don't really affect what I do. I'm not trying to optimize for growth. Unless the numbers dropped suddenly, I don't care too much about them. The feed is my own thing that I'm happy to share.''} (C4)
Creators largely treat these analytics as passive evaluations rather than metrics to optimize for.

\item \textbf{\textit{Receiving explicit feedback}}. Because there currently exists no direct way for consumers to communicate with the feed creator through the feed itself, creators actively solicit feedback from workarounds such as through commenting under their pinned posts or through auxiliary channels (\eg direct messaging, Discord, feedback forms). 
Creators reported acting on much of this feedback to improve their feeds, but emphasized that final decisions remain at their personal discretion.
% When creators do receive \textit{explicit} consumer and contributor feedback, they had made fair points and suggestions that creators acted upon to improve their feed. Creators caveat that while they do listen to feedback, their final decision on whether to act is fully up to them. 
For example, C19 reflected on balancing community input against their own judgment: \textit{``People might get offended that I allow posts about Israel bombing hospitals. My argument is that this is medically relevant, and the suppression of that kind of speech is what made me leave other platforms.''}
\end{itemize}

\subsection{Sustaining Custom Feeds}

We find that the task of sustaining a custom feed over time carries real costs in time, motivation, and money.

\subsubsection{\textbf{Creators' sustaining motivations are directly from hobbyism}}
Utility-providing creators tied their sustained motivation to a feed's continued growth. Stagnant feeds tended to go unmaintained—C21 reflected that while starting their feed was exciting, keeping it going without visible momentum proved difficult.
% \textit{``It's been really stagnant, which makes me sad. On Twitter, it felt like we had users at all levels. With the demise of Twitter, everyone scattered. I don't know how to attract more users to Bluesky.''}
On the other hand, \textit{community-building} creators are largely willing to sustain their feeds through ongoing personal investment, mirroring those of hobbyist moderators on platforms like Reddit and Discord~\cite{weld2025perceptions, seering2022metaphors}. 
C23 framed this volunteer labor as central to Bluesky's value: 
\begin{quote}
\textit{``I think if they go away from hobbyists, they're going to alienate the entire base that was cheering for Bluesky. I think it's a quick way to do the ``Vine upset,'' where you build this community and then crash it because you don't give back to the people that helped build it.''} \hfill --C23
\end{quote}

Creators were divided, however, on whether Bluesky itself adequately supports this volunteer labor.
Some creators viewed the current hobbyist model well-aligned with the decentralization philosophy of Bluesky. As C16 acknowledged,\textit{``I feel like Bluesky absolutely could do more. But I'm also on board with their mission for people to build their own things.''}
Others, like C11, found the absence of concrete platform support alienating, since the motivation to sustain a feed had to come entirely from the creator rather than from the platform:
% On the other hand, some creators lean strongly into the sentiment that Bluesky does not offer any concrete supports. C11 describes this as alienating, since the motivation to sustain their feeds must come from themselves rather than supported by Bluesky:
\textit{``I don't really feel any particular support from Bluesky. Maybe \textit{SkyFeed} feels differently, or someone who set up their own server, but nobody from Bluesky ever reached out or did anything to help me.''}

\subsubsection{\textbf{Creators distinguish personal monetization from ecosystem support.}}

Most third party tools are free to use, and many feeds' infrastructural needs are  able to be fully supported by free features. However, some tools have monetized features that creators desire to be monetarily supported. For example, \textit{BFC} follows a freemium model: basic feed setup is free, while a subscription unlocks more advanced capabilities such as additional post-capturing logic and translation~\cite{blueskyfeedcreator}.
Every 100 posts in a feed, \textit{BFC} also inserts an ad that this feed was created with \textit{BFC} which feed creators cannot remove. To cover the cost of premium tools like \textit{BFC}, creators solicit community support through donation links (\eg Ko-Fi), though creators state they do not personally take any of these funds and only use this to cover the cost of using a third party tool for the feed (C1, C6, C21).

Adjacently, \textit{Graze} has introduced an advertisement marketplace where feed creators can monetize their feeds, with potential advertisers placing an ad on the feed set at the creator's price~\cite{graze-marketplace}. 
Creators were divided on whether they should take advantage of it.
Many felt that personal monetization should not be necessary for sustaining their feeds, particularly when doing so would profit off content they do not own (\eg artwork).
As C16 explained, \textit{``People who are going to monetize are going to need to bring something very special and not easily replicated. If your feed is just a simple string match for hashtags, like some of mine, people will just build the same feed themselves, ad-free.''}.
These creators nonetheless accepted monetization that supported the infrastructure behind feeds.
C21, for instance, cited paying for \textit{BFC} and framed monetization in terms of cost recovery rather than profit: \textit{``It's not about getting rich; it's about covering the expense so I'm not paying out of pocket to help the community.''}

% However, creators were not against monetarily supporting the infrastructure that supports custom feeds. When it came to paying for a third party tool service or for supporting Bluesky as a platform, creators suggested that this should be inherently necessary. C21 felt, pessimistically, that the trend of monetization is eventually \textit{``inevitable''} for social media platforms. They cited that because \textit{BFC} is a paid third party tool, they would want support for this: \textit{``It’s not about getting rich; it’s about covering the expense so I’m not paying out of pocket to help the community.''}

A second group of creators went further, arguing that modest personal compensation can itself align with Bluesky's ethos. C20, who maintains a community-building art feed, framed feed creator monetization as part of a positive community loop:
\textit{``My feeling is that the person building and maintaining that feed has a right to ask for a couple of bucks. The people taking out ads on Bluesky are indie creators, authors, and self-publishing people. It's not Jeff Bezos; it's people giving \$20 to place an ad. That's supporting the creators, the feed builders, and the community. It's creating a positive loop of dollars going through the community.''}

C15's \textit{utility-providing} feed offers as a working example of this vision of monetization. This high-traffic news feed, one of the largest on Bluesky, has grown to sufficient scale to make advertising viable. 
All advertisers go through \textit{Graze}, and C15 personally approves every ad, with ads making up roughly 2\% of posts in the feed.
Each ad must also carry a specific hashtag for regulatory compliance, which doubles as a mechanism for consumers to opt out: \textit{``I've always been transparent about opting out. Users can just mute that hashtag if they want.''}
C15 estimates earning \$2,000--\$3,000 USD through \textit{Graze} over the past year and donates roughly half of these funds back to \textsc{ATProto} development projects.
They frame monetization not as personal income but as a mechanism for returning value to the ecosystem, putting C20's positive community loop into practice.

In contrast, some third party tools are simply not monetarily sustainable. For example, D2's tool hosts thousands of feeds on the ecosystem. Their tool has grown so large that it has become unsustainable to cover server hosting costs:
\begin{quote}
    \textit{``For the past two years, I've used GitHub Sponsors, and about 150 users have supported the project. This used to cover most of the server costs. I also received a \$1,000 grant from Bluesky once, which covered a couple of months. However, my primary goal has been to keep the service online as a hobby project, not to make money personally. Now, with the surge in growth around the U.S. elections, the server costs have become much more expensive, and donations no longer cover them. This is why I'm currently thinking about how to make the project sustainable, which might involve making it a full-time job and finding a way to pay myself in addition to the infrastructure costs.''} \hfill --D2
\end{quote}

%% file: figures/findings_figure.tex
\begin{figure}[t]
  \centering
  \includegraphics[width=1.0\textwidth]{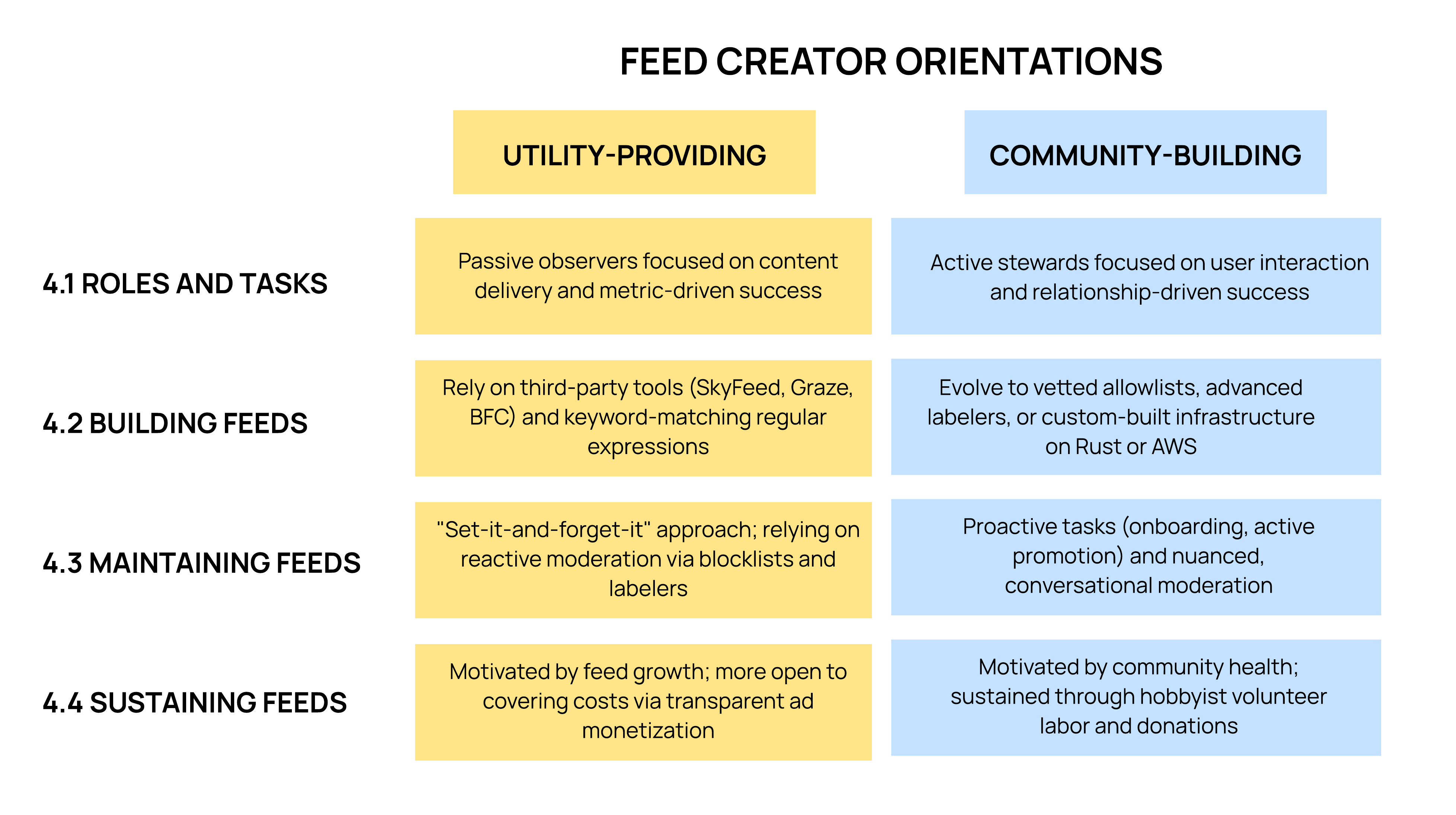}
  \caption{An overview of our findings mapped along our two feed creator orientations. Both utility-providing and community-building feed creators show distinctions across how they understand their roles and build, maintain, and sustain their custom feeds.}
  \label{figures:findings_figure}
\end{figure}

%% file: figures/active_consumers.tex
\begin{figure}[t]
  \centering
  \includegraphics[width=0.8\textwidth]{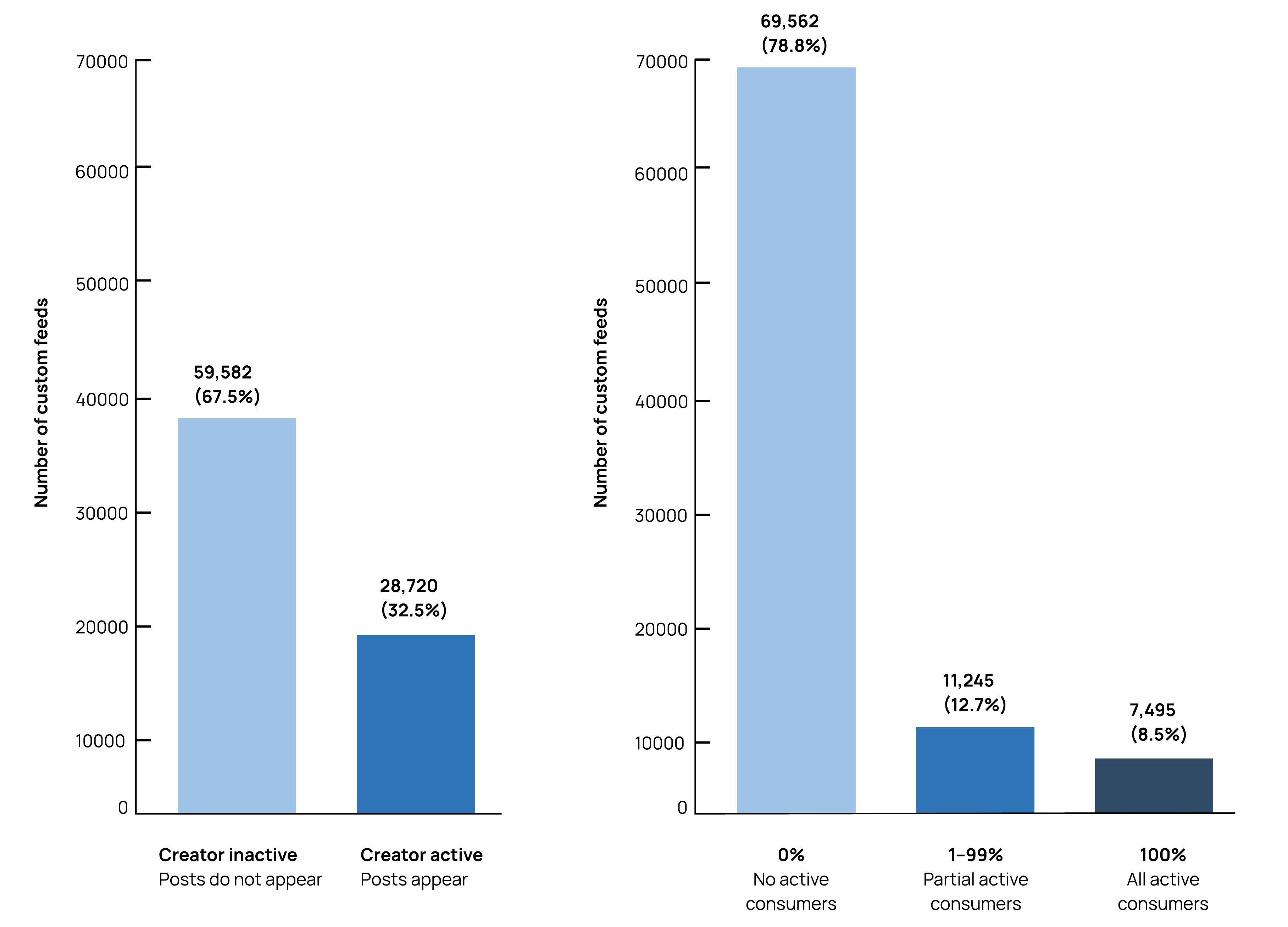}
  \caption{Proportions of creator and consumer engagement across active custom feeds in our dataset ($n=88,302$). The left histogram illustrates the frequency of creators posting within their own feeds, while the right histogram displays the distribution of active consumer participation rates.}
  \label{active_consumers}
\end{figure}

%% file: figures/interactions_engagement.tex
\begin{figure}[t]
  \centering
  \includegraphics[width=0.8\textwidth]{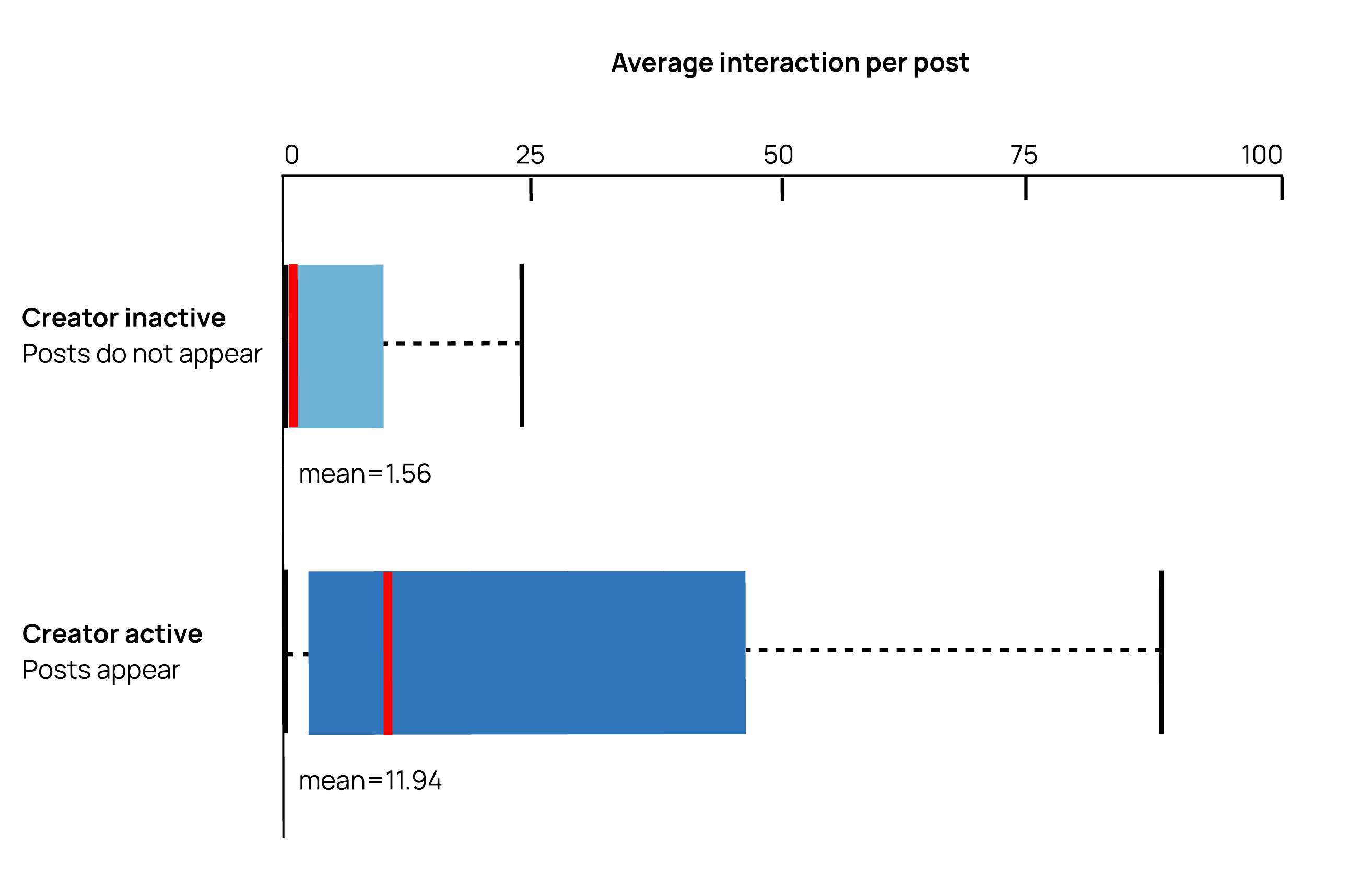}
  \caption{Boxplots of average interactions per post compared between feeds with inactive versus active feed creators. Feeds with active creators exhibit significantly higher average interactions per post. We omitted outliers above the 95th percentile for visual clarity.}
  \label{interactions_engagement}
\end{figure}

%% file: figures/third-party-tools.tex
\setlength{\tabcolsep}{5pt}
\renewcommand{\arraystretch}{1.25}
\begin{table}[h]
\centering
\small
\caption{Overview of the most popular third party tools used by feed creators on Bluesky.}
\begin{tabular}{p{2.8cm} p{2.7cm} p{2.7cm} p{3.3cm}}
\toprule
\textbf{Feature} &
\textbf{\textit{SkyFeed}} &
\textbf{\textit{Graze}} &
\textbf{\textit{Bluesky Feed Creator (BFC)}} \\
\midrule

\textbf{Feed logic}
&
Rule-based
&
Rule-based + ML
&
Rule-based
\\

\textbf{Moderation}
&
Regex filters
&
Block lists, labelers
&
Block lists, labels
\\

\textbf{Creator analytics}
&
Basic
&
Advanced
&
Basic
\\

\textbf{Monetization}
&
None
&
Ads
&
Donations
\\

\bottomrule
\end{tabular}
\label{tab:thirdpartytools}
\end{table}

%% file: sections/06-discussion.tex
\section{Discussion}

\subsection{Decentralized Middleware in Practice: Power and Cost Asymmetries}

The custom feed ecosystem offers a large-scale example of the middleware vision for algorithmic feed recommendation, giving actors beyond the Bluesky platform (\eg third-party tool developers, feed creators, and consumers) direct agency in shaping how feeds are designed and maintained (see Figure ~\ref{figures:findings_figure}). At the same time, our findings show that this ecosystem contains important asymmetries across these actors, especially in who \textit{holds control} and who \textit{bears the costs} of sustaining custom feeds.

\subsubsection{{Power asymmetries in platform and tooling control}}
Bluesky still retains several forms of power that remain inaccessible to feed creators and third-party developers. The most direct is privileged access to user data, since 98.9\% of users' data in 2024 were still hosted on Bluesky's own servers ( \texttt{bsky.social}) rather than on self-hosted alternatives~\cite{balduf2024looking}. As D2 explained,\textit{``The official \textsc{Discover} feed from Bluesky has a special advantage because it can collect a ton of interaction metrics (clicks, views, etc.) that third-party developers can't.''}
Bluesky also exercises power through less visible interface choices, shaping how accessible custom feeds are to users. For instance, custom feeds are not prominently surfaced during onboarding or in the platform's main UI (see Figure ~\ref{fig:main-figure}) and creators are frustrated with the lack of support from Bluesky itself to promote feed discovery.

% to support feed discovery Bluesky's built-in feed discovery support is widely seen as ineffective. 
% why is it seen as ineffective? introduce this earlier

At the tooling layer, power concentrates around a small number of third-party developers whose design choices determine what features feed creators can use. The deeper asymmetry, however, stems from technical capability itself. Only three interviewed feed creators had the expertise to bypass third-party tooling entirely and build their own infrastructure. Most others struggled with even the regex-based filters that underlie these tools. Due to these power asymmetries at both the platform and tool levels, many creators were frustrated that their feeds would never be as popular or performant as Bluesky's \textsc{Discover} feed despite the ecosystem’s formally open architecture for custom feeds. {As we were finishing this paper, a personalized \textit{For You} feed\footnote{https://bsky.app/profile/spacecowboy17.bsky.social/feed/for-you} became the most liked custom feed on the platform, suggesting that within the ostensibly open ecosystem, the most popular feeds still reproduce prior centralized, personalized algorithms.}

%TODO look into
%myth or the harm of "technical expertise" or basically determining who and who isn't considered a technologist based on their traditional technical skills. I would maybe dive into this literature to get more insight on how the frame the technical power imbalance.

\subsubsection{{Cost asymmetries in sustaining custom feeds}}

Beyond power asymmetries, the custom feed ecosystem is sustained primarily through hobbyist labor, paralleling prior work on volunteer moderators on Reddit and Discord~\cite{seering2022metaphors, weld2025perceptions}. In contrast to traditional content moderators, feed creators also must take on extensive additional work introduced by this new role, such as gathering user feedback and promoting their feeds, with, unfortunately, little platform support. 
Beyond time commitments, sustaining the ecosystem incurs financial costs as well, most notably the servers required to host curation algorithms and store user data. Third-party tools such as \textit{BFC} shift some of these costs onto creators through subscription fees for advanced features. Creators either absorb these costs themselves or recover them through donations and, increasingly, in-feed advertising.
While these efforts represent an early attempt to promote financial sustainability of this ecosystem, our interviews surfaced several unresolved tensions, such as whether feed creators are worth compensating for their personal efforts beyond infrastructure cost, and, furthermore, whether in-feed ads {should} be allowed in custom feeds even if they can align with the ethos of Bluesky.

% TODO mention enshitification? 
\subsubsection{\textbf{Implications}} 

Our findings suggest that improving and sustaining the custom feed ecosystem requires directly addressing the power and cost asymmetries that currently burden creators and consumers.
Meaningful decentralization requires reducing both the power gap (concentrated data and tooling control) and the cost gap (disproportionate creator burdens), while avoiding interventions that would create new forms of centralization.

\begin{enumerate}
    \item \textbf{Future decentralized infrastructures could separate data portability from data visibility.} Interaction signals remain exclusively accessible to Bluesky's \textsc{Discover} feed, giving platform-operated feeds a structural advantage that third-party creators cannot close through feed logic alone. While Bluesky is alleviating this by starting to provide data to developers (as C2 mentioned), we posit that Bluesky could offer aggregated anonymized engagement data (\eg which posts in a feed were clicked versus scrolled past) without exposing individual user behavior. Furthermore, there could be features in which feed subscribers voluntarily label posts and provide feedback on whether feeds are matching individual and collective preferences. These possibilities could provide data for creators to improve their feeds that does not infringe upon users' privacy.

    %i would perhaps mention designs for more accessible feed creation assisted by AI. for instance, having a conversation with AI to elicit how feed creators want to build feeds similar to the social media elicitation paper, or having AI to author these regular expression or more advanced features. 
    \item \textbf{AI-assisted features could significantly lower barriers to the technical complexity of feed creation.} Researchers have found that conversational elicitation could help individual users articulate what content they want to surface or suppress for their custom feeds, and AI can assist with translating preferences into regular expressions or more advanced feed logic~\cite{malki2025bonsai, popowski2026social}. We envision that such systems could similarly help feed creators author their custom feeds, and help feed consumers articulate their feedback more accurately. For instance, a tool could surface examples of content that a feed would include and exclude and verify with creators whether these match their intent. This would expand the design space for non-technical creators beyond what current low- or no-code tools support.

    \item \textbf{Sustaining the custom feed ecosystem requires supporting feed discovery and creator maintenance work.} 
    Our findings suggest that feed discovery could be supported more directly by platform infrastructure, reducing the need for creators to self-promote. Existing third party websites\footnote{https://www.bskyinfo.com/feeds/} aggregate feeds into topical lists for ease of search, but it is unclear how effective this is for individual users to discover feeds. A potential solution could be Bluesky utilizing a simple personalized ranking algorithm to recommend feeds based on user interests through their already public data as a complement to the \textsc{Discover} feed. But we note that this can raise the question of who designs that ranking algorithm and by what criteria, reintroducing another form of centralized power over which feeds gain visibility. Additionally, we believe collective costs can be sustained through community-guided labeler, moderation, and monetization systems for creators to lower the operational burden of maintenance.
    
\end{enumerate}

% input figure

% why do we need to do this?
% TODO -> can we enrich this a bit more? 
% why did we do this? 
%  
\subsection{Characterizing the Feed Landscape Through the Lens of Powers}

End users today choose among a broad and growing landscape of feeds: commercial personalized feeds curated by platforms (\eg TikTok's \textit{For You} page), approaches that let individuals assemble feeds for their own consumption (\eg Bonsai~\cite{malki2025bonsai}, feed elicitation interviews~\cite{popowski2026social}, Compass~\cite{barua2026compass}), creator-built custom feeds like those on Bluesky, and established community platforms such as Reddit subreddits. These feeds differ less in the content they carry than in who is \textit{entitled to shape them, and how}.
{Extending prior work that discerns platform power dynamics in decentralized social media protocols~\cite{oshinowo2025seeing}, we reflect on the feeds in our study, and characterize this landscape along three forms of power over a feed (see Table ~\ref{fig:rights_framework})}:

\begin{itemize}
    \item \textbf{The power to host}: control over the infrastructure a feed runs on, \eg the servers that execute its logic and store its data.
    \item \textbf{The power to curate}: authority over a feed's selection and ranking logic---what content it surfaces. This power also depends on what data the actor has access to, and their engineering resources. A platform that observes every click and dwell time can curate in ways that a creator limited to public posts cannot.
    \item \textbf{The power to contribute}: the ability to push content directly into a feed that others consume. Unlike the first two, this right is graded: a news homepage gives readers no way to contribute; feeds like TikTok's \textit{For You} page let anyone post but leave the odds of surfacing to the platform; and some feeds let contributors place content straight into what consumers see.
\end{itemize}

Where these powers sit distinguishes one feed from another, and further shapes what consumers experience.
A commercial personalized feed like TikTok's \textit{For You} keeps all three with the platform: it hosts the feed, sets its logic, and tightly controls what users can contribute. 
This full power lets the platform build more effective recommendation feeds than any other actor can.
Custom feeds, by contrast, pull these rights apart. \textit{Utility-providing} feeds delegate hosting to a third-party tool while the creator holds curation, though consumers typically have no direct power to contribute to the feed. \textit{Community-building} feeds extend \textit{direct} contribution to consumers themselves. Personal feed-building tools, in turn, hand curation to individual users, who define a feed's logic for their own consumption. Across these forms, custom feeds occupy a distinctive region of this spectrum, where the three powers rarely rest with a single actor. 
This decentralization comes at the cost of recommendation quality, but closing that gap would require the pervasive data collection that motivated many users to leave prior platforms in the first place. 
Viewed through this lens, the boundary between a \textit{feed} and an \textit{online community} itself begins to blur: a \textit{community-building} feed, for instance, becomes hard to cleanly separate from a subreddit.

\input{figures/rights_framework}

\subsection{Holding Feed Creators Accountable: Transparency of Feed Logic and Its Trade-off}

% While improving recommendation quality is crucial for the success of the custom feed ecosystem, this challenge may potentially introduce many aspects of centralization that motivated Bluesky users to leave prior platforms. Third-party tools lack access to implicit interaction signals (\eg clicks, dwell time, passive engagement) and leveraging these signals requires sophisticated algorithms to determine behavioral patterns, which users may not be comfortable with.

% explicitly connect with above? 

As custom feeds move the power to curate from platforms to feed creators, that power might now be exercised with far less oversight than platforms face. Some creators we interviewed worried that their feeds could easily be repurposed or acquired without subscribers knowing. C14 illustrated the risk by converting a joke feed into one about trans rights overnight while dozens of subscribers still had it pinned. Similarly, C16 worried that \textit{``I can come to you and say, `Hey, I'll buy the feed off you,' and no one would ever know. Then I could pivot it to my own content, my own agenda. It's the same as when people sell their large Reddit accounts so certain opinions can get more traction.''} Feed creators may also come to hold more of consumers' interaction data than consumers would willingly share, especially as Bluesky has recently begun releasing fine-grained interaction data to feed developers (C2). D2 was wary of this direction: \textit{``I would prefer if this data wasn't collected at all, as I dislike multiple entities being informed of my actions. I worry it will be used for personalized ads in the future.''} 

While the middleware ecosystem lets consumers keep their data and switch to a different feed, a more direct way to hold creators accountable for this power is to make their feed logic transparent to consumers.
A central design question, then, is how much of a feed's logic should be exposed to consumers. Today, consumers can only infer what a feed does from its description and its content; some third party tools publish the full definition of every feed they host, but most feeds expose nothing. Publishing feed logic  would let consumers judge whether a feed still matches what they subscribed to and hold creators to their stated purpose. However, such transparency also invites abuse of the power to contribute: once contributors know the exact keywords a feed matches, they can insert them to push their own content into it, as creators already experience with hashtag spam. To address this trade-off, we envision exposing what a feed does rather than how it does it, for instance by showing sample posts it surfaces and suppresses.

%% file: figures/rights_framework.tex
\setlength{\tabcolsep}{5pt}
\renewcommand{\arraystretch}{1.25}
\begin{table}[h]
\centering
\small
\caption{A comparison of feed curation paradigms across  three rights of feed governance: the right to host, the right to curate, and the right to contribute.}
\begin{tabular}{p{2.8cm} p{2.7cm} p{2.7cm} p{3.3cm}}
\toprule
\textbf{Feed Paradigm} &
\textbf{Power to Host} &
\textbf{Power to Curate} &
\textbf{Power to Contribute} \\
\midrule

\textbf{Commercial feeds} \newline (\eg TikTok, Instagram)
&
Centralized platform
&
Centralized platform
&
Algorithmic gatekeeping by platform
\\

\textbf{Custom feeds by \newline utility-providing \newline creators} \newline (\eg C9, C15)
&
Third-party tool \newline or self-hosted
&
Feed creator
&
Read-only for consumers
\\

\textbf{Custom feeds by \newline community-building \newline creators} \newline (\eg C14, C19)
&
Third-party tool \newline or self-hosted
&
Feed creator
&
Direct contribution (\eg via hashtags or replies)
\\

\textbf{Personal feed tooling} \newline (\eg Gobo~\cite{bhargava2019gobo}, Bonsai~\cite{malki2025bonsai}), Compass~\cite{barua2026compass}
&
Client-side \newline or self-hosted
&
Individual consumer
&
Limited to a user's personal scope
\\

\bottomrule
\end{tabular}
\label{fig:rights_framework}
\end{table}

%% file: sections/07-limitations.tex
\section{Limitations and Future Work}
%user perspective
%participant perspective may not be fully representative

% We used feed likes as a proxy of engagement as a result, but we note that users can still view feeds without having liked or pinned a feed.

We recognize two limitations with our work. First, we did not directly solicit perspectives from feed consumers or contributors. This is because we hypothesized that feed creators' experiences would be a more representative view of the custom feed ecosystem. Existing third party tools are built for supporting feed creators and feed creators are publicly contactable from both the Bluesky web application and public lists of custom feeds, which supported our recruitment for creators rather than contributors or consumers. Our findings also highlight that communicating with consumers is difficult even for feed creators themselves. A promising future direction could focus on understanding {consumer} perspectives of custom feeds and algorithmic curation on Bluesky.

Second, our quantitative analysis does not aim to classify custom feed logic or algorithms at scale. While we attempted to run an LLM-based inference to classify custom feed logic based on feed descriptions and the last 500 posts in a feed, we could not draw meaningful conclusions from it. Many unkept custom feeds do not have descriptions to draw from or whose algorithm does not show many recent posts, and the vast majority of feeds still hold little to no indication as to how its' algorithm functions. As Bluesky grows and draws more interest in research, we encourage future work to quantitatively categorize custom feed algorithms and logic at scale.

% This is partially due to Bluesky not providing any API to access a specific feed's logic. 

% Additionally, we recognize that our interview participant pool may not be fully representative of all feed creators on Bluesky. Future work should consider understanding the needs of all populations on Bluesky.

%% file: sections/08-conclusion.tex
\section{Conclusion}

Custom feeds on Bluesky represent the first large-scale instantiation of a long-envisioned \textit{middleware}-centered feed recommendation model for social media. By returning meaningful control over algorithmic curation to users and third-party developers, the growing custom feed ecosystem demonstrates that middleware can exist in practice for feed recommendation at scale, allowing users of all kinds to achieve new control over their online experiences beyond simply centralized algorithmic recommendations. Feed creators have built a diverse range of algorithmic infrastructure ranging from simple keyword filters to large-scale community spaces, sustained largely through hobbyist labor and third-party tooling instead of direct platform support.

However, our findings show that these middleware ideals remain only partially realized in practice. We find that power and infrastructural labor remain unevenly distributed across actors in the ecosystem, particularly stressed on third party developers and feed creator labor. This is exacerbated by gaps between what Bluesky provides for third party tool developers, the normalized roles of creators, and also compounded by tensions between personalization, transparency, and community-oriented curation. We enumerate future guidelines to support these visions, and we hope our work provides a foundation for future research and design toward more usable and sustainable decentralized social media ecosystems.

%% file: sections/appendix.tex
\appendix
\section{Participant Demographics}
\label{appendix:participants}

%>1000 n=8
%500-999 n=9
%100-499 n=5
%1-99 n=4
%mention the scale of their custom feeds either in the recruitment paragraph or here. 
%and also smaller feeds. 
% Our recruitment criteria ensured that:

% \begin{itemize}
%     \item Participants were above the age of 18.
%     \item Participants had experience creating at least one custom feed on Bluesky, or a third party tool.
%     \item Participants had self-reported a basic level of expertise or higher with creating custom feeds.
% \end{itemize}

\begin{table*}[h]
  \centering
  \small
  \begin{tabular}{ l l l l l l l r }
  \toprule
  \textbf{ID} & \textbf{Age} & \textbf{Gender} & \textbf{Country} & \textbf{Social Media Use} & \textbf{Feed Expertise} \\
  \midrule
  \textbf{C1} & 35--44 & Man & Australia & 5--9 & Expert\\
  \textbf{C2} & 35--44 & Man & USA & 10+ & Expert\\
  \textbf{C3} & 18--24 & Woman & USA & 10+ & Competent\\
  \textbf{C4} & 25--34 & Prefer not to say & USA & 5--9 & Competent\\
  \textbf{C5} & 25--34 & Woman & USA & 1--2 & Expert \\
  \textbf{C6} & 25--34 & Man & USA & 10+ & Expert \\
  \textbf{C7} & 35--44 & Man & USA & 10+ & Competent\\
  \textbf{C8} & 45--54 & Man & Canada & 5--9 & Expert\\
  \textbf{C9} & 25--34 & Man &  USA & 10+ & Expert\\
  \textbf{C10} & 18--24 & Man & Germany & 10+ & Expert\\
  \textbf{C11} & 45-54 & Man & USA & 10+ & Expert\\
  \textbf{C12} & 55-64 & Man & USA & 10+ & Competent\\
  \textbf{C13} & 35-44 & Woman & USA & 10+ & Competent\\
  \textbf{C14} & 25-34 & Non-binary & Brazil & 10+ & Competent\\
  \textbf{C15} & 35-44 & Man & USA & 10+ & Competent\\
  \textbf{C16} & 35-44 & Non-binary & USA & 10+ & Expert\\
  \textbf{C17} & 35-44 & Non-binary & UK & 10+ & Expert\\
  \textbf{C16} & 35-44 & Man & USA & 10+ & Competent\\
  \textbf{C17} & 25-34 & Man & USA & 5--9 & Competent\\
  \textbf{C18} & 35-44 & Woman & USA & 10+ & Beginner\\
  \textbf{C19} & 25-34 & Man & USA & 10+ & Expert\\
  \textbf{C20} & 35-44 & Woman & USA & 10+ & Competent\\
  \textbf{C21} & 45-54 & Prefer not to say & Canada & 10+ & Beginner \\
  \textbf{C22} & 45-54 & Woman & USA & 10+ & Beginner\\
  \textbf{C23} & 25-34 & Woman & Canada & 10+ & Beginner\\
  \textbf{C24} & 35-44 & Woman & USA & 10+ & Expert\\

  % 1-50; 50-100; 100-500; 500+? 
  
  \bottomrule
  \end{tabular}
  \caption{Study participant demographics. Participants were asked to provide their years of social media usage and rate their custom feed expertise on a four-point scale (Novice: Has no or minimal knowledge, Beginner: Basic knowledge of custom feeds, Competent: Good working or background knowledge, or Expert: Has in-depth knowledge and understanding).}
\end{table*}

\newpage

%\caption{Study participant demographics collected through our sign-up form. Participants were asked to rate their custom feed expertise on a four-point scale (Novice: Has no or minimal knowledge, Beginner: Basic knowledge of custom feeds, Competent: Good working or background knowledge, or Expert: Has in-depth knowledge and understanding).}

% \newpage

\section{Study Information}
\label{appendix:interviews}

\subsection{Interview Protocol}

% Please note that this is the protocol for our second round of interviews, which we iterated on from our first round.

\subsubsection{Motivations}
\begin{itemize}
    \item What inspired you to start using Bluesky as a social media platform?
    \item How long after joining did you consider making your own custom feeds?
    \item What kinds of custom feeds have you created on Bluesky?
    \begin{itemize}
        \item What are their topics? 
        \item How many feeds did you create and how long did it take?
        \item How did you initially start creating them?
        \item Did you use any third party feed-building tools? If so, how was that experience?
        \item Why did you decide to create a custom feed in the first place?
    \end{itemize}

    \item How long have you still been actively maintaining your first feed? Or any others?

    \item What motivates you to still keep maintaining your feed?

\end{itemize}

\subsubsection{Feed Metrics and Growth}

\begin{itemize}
    \item How many subscribers have your feeds attracted?
    \item How many posts does your feed publish every day or every week?
    \item Do you keep track of any specific feed metrics, such as interactions, engagement, or click-throughs?
    \begin{itemize}
        \item If so, how does tracking these metrics help you understand growth or plan for the future?
        \item Which metrics matter most to you?
        \item Given all these metrics you track, what actually makes you feel like your feed is working or successful?
    \end{itemize}
    \item Have you been inspired to create more feeds after one took off?
\end{itemize}

\subsubsection{Maintenance Tasks}

\begin{itemize}
    \item Can you walk me through what a typical week looks like maintaining your feed? 
    \begin{itemize}
        \item What tasks are you working on, and what improvements are you thinking about?
        \item How easy or enjoyable is it for you? 
    \end{itemize}
    \item Do you feel you can keep dedicating this time and effort to your feed? Do you wish it were more or less? 
    \item Do you see yourself doing this a year from now? Two years? Five years?
\end{itemize}

\subsubsection{Moderation and Content Curation}

\begin{itemize}
    \item What kinds of unwanted content or behavior occurs on your feed? (spam, off-topic content, incivility, etc.)
    \begin{itemize}
        \item How do you find this type of unwanted content?
        \item Do your subscribers ever report content to you?
        \item How often do you patrol or monitor your feed? 
    \end{itemize}
    \item How do you deal with removing unwanted content from your feeds? 
        \begin{itemize}
            \item Do you use third party tools? Does Bluesky provide any support? 
            \item Do you then update your feed rules accordingly? 	
            \item Have you ever encountered a case where someone tried to exploit your algorithm rules to get into your feed? 
            \item Do you ever consider moderating comments or interactions by users that have been viewed through your feed? 
            \item What is your perceived role in this kind of moderation? 
            \item Do you feel a sense of achievement or happiness curating feeds for your audience? 
            \item Are there moments of frustration in this process? 
        \end{itemize}
    \item How do you communicate your feed rules or guidelines to your subscribers?
    \begin{itemize}
        \item How transparently are your content curation rules communicated to your subscribers? 
        \item How do you balance the tradeoff and potential risks between communicating more or less of this? 
    \end{itemize}
\end{itemize}

% \subsection{Web-Hosted Recruitment Message}

% \subsubsection{What is this project about?}

% For decades, centralized social media platforms have controlled what users see through opaque, top-down algorithms. While today these algorithms have grown to excel at generating engagement based on large-scale behavioral data and predictive models, users have little say in what content appears in their feeds or prioritizing the content that they want to see.

% Bluesky's custom feeds represent a fundamental shift to this approach, providing a unique opportunity for any user on the platform to create, share, and subscribe to their own feed algorithms. This opens up exciting possibilities for user agency and community-driven content.

% As this ecosystem of new feeds grows, we're particularly interested in understanding:

% \begin{itemize}
%     \item How are feed creators building and maintaining feeds as they grow? What practices emerge and what challenges arise?
%     \item What motivates creators to sustain their feeds, beyond hobbyist experiments into thriving communities?
%     \item How might custom feeds become a fully viable alternative to centralized recommendation systems: and what would it take to get there?
% \end{itemize}

% Your experiences as a feed creator are pioneering new models for social media. By understanding what works, what's challenging, and what support you need, we hope to inform better tools and contribute to the broader vision of decentralized content curation!

\newpage
\section{Feed Collection Data}

A sample JSON object we collected for each custom feed.
%numbers=none
\begin{lstlisting}[
    language=python,
    basicstyle=\ttfamily\small,
    showstringspaces=false,     
    breaklines=true,
    breakatwhitespace=false,
    frame=single,
    columns=fullflexible,      
    keepspaces=true           
]
  {
    "uri": "at://did:plc:z72i7hdynmk6r22z27h6tvur/app.bsky.feed.generator/thevids",
    "cid": "bafyreigsetw3cy2ryav7lwuzv7kwhh5ay7apnruva2ybsorsee7efzn4aa",
    "did": "did:web:discover.bsky.app",
    "creator_did": "did:plc:z72i7hdynmk6r22z27h6tvur",
    "creator_handle": "bsky.app",
    "display_name": "Video",
    "description": "Trending videos in the Bluesky network",
    "avatar": "https://cdn.bsky.app/img/avatar/plain/did:plc:z72i7hdynmk6r22z27h6tvur/bafkreicrnrylrdlwja5rsku7763xw6o2bjopkhvl3rubpaf4t54lkss2ja@jpeg",
    "like_count": 6710,
    "indexed_at": "2025-01-19T20:07:12.948Z",
    "collected_at": "2026-01-18T13:08:07.952667"
  },
\end{lstlisting}

\label{appendix:json}

\newpage

\section{Descriptive statistics of our dataset}
\label{appendix:descriptive_statistics}
\input{figures/descriptive_statistics}

%% file: figures/descriptive_statistics.tex
\begin{table}[h]
\centering
\scriptsize
\caption{Descriptive statistics of our custom feed dataset.}
\label{tab:descriptive_statistics}
\begin{tabular}{lrr}
\toprule
\textbf{Metric} & \textbf{Value} \\
\midrule
\multicolumn{2}{l}{\textit{Collection overview}} \\
\quad Total feeds collected & 88,302 \\
\quad Feeds with $\geq$1 subscriber & 49,069 (55.6\%) \\
\quad Unique feed creators & 41,955 \\
\quad Total posts scanned & 3,695,724 \\
\quad Total unique subscribers & 1,319,921 \\
\midrule
\multicolumn{2}{l}{\textit{Subscribers per feed} (feeds with $\geq$1 subscriber)} \\
\quad Median & 3 \\
\quad Mean & 26.9 \\
\quad Maximum & 50,757 \\
\quad 1--10 subscribers & 37,801 (77.0\%) \\
\quad 11--100 subscribers & 9,908 (20.2\%) \\
\quad 101--1,000 subscribers & 1,239 (2.5\%) \\
\quad 1,000+ subscribers & 121 (0.2\%) \\
\midrule
\multicolumn{2}{l}{\textit{Posts per feed}} \\
\quad Median posts scanned & 18 \\
\quad Mean posts scanned & 41.9 \\
\quad Feeds with 0 posts & 21,517 (24.4\%) \\
% \quad Feeds reaching 500-post window & 1,826 (2.1\%) \\
\midrule
\multicolumn{2}{l}{\textit{Feed age}} \\
\quad Median & 612 days (1.7 yrs) \\
\quad Mean & 631 days (1.7 yrs) \\
\midrule
\multicolumn{2}{l}{\textit{Content composition} (median / mean)} \\
\quad Media ratio & 75.0\% / 59.4\% \\
\quad Reply ratio & 0.0\% / 6.2\% \\
\quad Non-chronological ordering & \multicolumn{1}{r}{15.1\%} \\
\midrule
\multicolumn{2}{l}{\textit{Language} (feeds with $\geq$1 subscriber)} \\
\quad English & 25,981 (52.9\%) \\
\quad Unknown (no tag) & 11,638 (23.7\%) \\
\quad Japanese & 7,423 (15.1\%) \\
\quad German & 907 (1.8\%) \\
\quad Portuguese & 679 (1.4\%) \\
\bottomrule
\end{tabular}
\end{table}

%% file: 99-bib.bib
@article{liu2025understanding,
  title={Understanding decentralized social feed curation on mastodon},
  author={Liu, Yuhan and Song, Emmy and Zhang, Owen Xingjian and Merriman, Jewel and Zhang, Lei and Monroy-Hern{\'a}ndez, Andr{\'e}s},
  journal={Proceedings of the ACM on Human-Computer Interaction},
  volume={9},
  number={7},
  pages={1--25},
  year={2025},
  publisher={ACM New York, NY, USA}
}

@article{geiger2016bot,
  title={Bot-based collective blocklists in Twitter: the counterpublic moderation of harassment in a networked public space},
  author={Geiger, R Stuart},
  journal={Information, Communication \& Society},
  volume={19},
  number={6},
  pages={787--803},
  year={2016},
  publisher={Taylor \& Francis}
}

@article{zhang2024form,
  title={Form-from: A design space of social media systems},
  author={Zhang, Amy X and Bernstein, Michael S and Karger, David R and Ackerman, Mark S},
  journal={Proceedings of the ACM on Human-Computer Interaction},
  volume={8},
  number={CSCW1},
  pages={1--47},
  year={2024},
  publisher={ACM New York, NY, USA}
}

@inproceedings{sokoto2026open,
  title={Open or Blocked Skies? Community Moderation Practices in Bluesky},
  author={Sokoto, S and Badhuf, L and Ascigil, O and Tyson, G and Castro, I and Scheuermann, B and Baronchelli, A and Krol, M},
  booktitle={ACM The Web Conference},
  year={2026}
}

@article{poell2020three,
  title={Three challenges for media studies in the age of platforms},
  author={Poell, Thomas},
  journal={Television \& New Media},
  volume={21},
  number={6},
  pages={650--657},
  year={2020},
  publisher={Sage Publications Sage CA: Los Angeles, CA}
}

@inproceedings{atreja2023understanding,
  title={Understanding journalists’ workflows in news curation},
  author={Atreja, Shubham and Srinath, Shruthi and Jain, Mohit and Pal, Joyojeet},
  booktitle={Proceedings of the 2023 CHI Conference on Human Factors in Computing Systems},
  pages={1--13},
  year={2023}
}

@article{rashed2025if,
  title={What If Moderation Didn't Mean Suppression? A Case for Personalized Content Transformation},
  author={Rashed, Rayhan and Jahanbakhsh, Farnaz},
  journal={arXiv preprint arXiv:2509.22861},
  year={2025}
}

@inproceedings{henderson2025graffiti,
  title={Graffiti: Enabling an Ecosystem of Personalized and Interoperable Social Applications},
  author={Henderson, Theia and Karger, David R and Clark, David D},
  booktitle={Proceedings of the 38th Annual ACM Symposium on User Interface Software and Technology},
  pages={1--21},
  year={2025}
}

@inproceedings{boeker2022empirical,
  title={An empirical investigation of personalization factors on TikTok},
  author={Boeker, Maximilian and Urman, Aleksandra},
  booktitle={Proceedings of the ACM web conference 2022},
  pages={2298--2309},
  year={2022}
}

@inproceedings{backstrom2011supervised,
  title={Supervised random walks: predicting and recommending links in social networks},
  author={Backstrom, Lars and Leskovec, Jure},
  booktitle={Proceedings of the fourth ACM international conference on Web search and data mining},
  pages={635--644},
  year={2011}
}

@article{masnickprotocolsnotplatforms,
title={Protocols, Not Platforms: A Technological Approach to Free Speech},
author={Masnick, Mike},
journal={Knight First Amendment Institute},
year={2019},
note={https://knightcolumbia.org/content/protocols-not-platforms-a-technological-approach-to-free-speech}
}

@inproceedings{wang2025end,
  title={End user authoring of personalized content classifiers: Comparing example labeling, rule writing, and llm prompting},
  author={Wang, Leijie and Yurechko, Kathryn and Dani, Pranati and Chen, Quan Ze and Zhang, Amy X},
  booktitle={Proceedings of the 2025 CHI Conference on Human Factors in Computing Systems},
  pages={1--21},
  year={2025}
}

@book{wenger1999communities,
  title={Communities of practice},
  author={Wenger, Etienne},
  volume={92},
  year={1999},
  publisher={Cambridge university press Cambridge}
}

@incollection{boyd2010social,
  title={Social network sites as networked publics: Affordances, dynamics, and implications},
  author={Boyd, Danah},
  booktitle={A networked self},
  pages={47--66},
  year={2010},
  publisher={Routledge}
}

@article{berlant1998intimacy,
  title={Intimacy: A special issue},
  author={Berlant, Lauren},
  journal={Critical inquiry},
  volume={24},
  number={2},
  pages={281--288},
  year={1998},
  publisher={University of Chicago Press}
}

@inproceedings{fischer2001communities,
  title={Communities of interest: Learning through the interaction of multiple knowledge systems},
  author={Fischer, Gerhard},
  booktitle={Proceedings of the 24th IRIS Conference},
  volume={1},
  pages={1--13},
  year={2001},
  organization={Department of Information Science, Bergen}
}

@article{barua2026compass,
  title={Compass: Continuously Aligning Social Media Feeds via In-Situ Reflections},
  author={Barua, Aadit and Wang, Leijie and Zhang, Amy X},
  journal={arXiv preprint arXiv:2608.04274},
  year={2026}
}

@article{zuckerman2024improving,
  title={Improving social media with middleware},
  author={Zuckerman, Ethan and Brickman, Isaac},
  journal={The ANNALS of the American Academy of Political and Social Science},
  volume={715},
  number={1},
  pages={99--114},
  year={2024},
  publisher={SAGE Publications Sage CA: Los Angeles, CA}
}

@misc{blue-facts,
key = {Bluesky Top Feeds and Statistics
},
note = {https://bluefacts.app/feeds}
}

@misc{bluesky-history,
    key = {About Bluesky.},
    date = {2025},
    note = {https://bsky.social/about/faq}
}

@misc{atproto,
key = {AT PROTOCOL.
Building the Social Internet.},
date = {2026},
note = {https://atproto.com/}
}

@inproceedings{balduf2025bootstrapping,
  title={Bootstrapping Social Networks: Lessons from Bluesky Starter Packs},
  author={Balduf, Leonhard and Sokoto, Saidu and Baronchelli, Andrea and Castro, Ignacio and Kr{\'o}l, Micha{\l} and Tyson, Gareth and Pavlou, George and Scheuermann, Bj{\"o}rn and Ascigil, Onur},
  booktitle={Proceedings of the International AAAI Conference on Web and Social Media},
  volume={19},
  pages={178--192},
  year={2025}
}

@misc{bluesky-algorithmic-choice,
    author = {Bluesky Team},
    key = {Algorithmic Choice with Custom Feeds.},
    date = {2024},
    note = {https://bsky.social/about/blog/7-27-2023-custom-feeds}
}

@article{parnin2012crowd,
  title={Crowd documentation: Exploring the coverage and the dynamics of API discussions on Stack Overflow},
  author={Parnin, Chris and Treude, Christoph and Grammel, Lars and Storey, Margaret-Anne},
  year={2012}
}

@inproceedings{chan2026examining,
  title={Examining algorithmic curation on social media: An empirical audit of reddit’sr/popular feed},
  author={Chan, Jackie and Choi, Fred and Saha, Koustuv and Chandrasekharan, Eshwar},
  booktitle={Proceedings of the International AAAI Conference on Web and Social Media},
  volume={20},
  number={1},
  pages={391--406},
  year={2026}
}

@article{kleinberg2024challenge,
  title={The challenge of understanding what users want: Inconsistent preferences and engagement optimization},
  author={Kleinberg, Jon and Mullainathan, Sendhil and Raghavan, Manish},
  journal={Management science},
  volume={70},
  number={9},
  pages={6336--6355},
  year={2024},
  publisher={INFORMS}
}

@inproceedings{bhargava2019gobo,
  title={Gobo: A system for exploring user control of invisible algorithms in social media},
  author={Bhargava, Rahul and Chung, Anna and Gaikwad, Neil S and Hope, Alexis and Jen, Dennis and Rubinovitz, Jasmin and Sald{\'\i}as-Fuentes, Bel{\'e}n and Zuckerman, Ethan},
  booktitle={Companion publication of the 2019 conference on computer supported cooperative work and social computing},
  pages={151--155},
  year={2019}
}

@article{choi2025designing,
  title={Designing Usable Controls for Customizable Social Media Feeds},
  author={Choi, Frederick and Chandrasekharan, Eshwar},
  journal={arXiv preprint arXiv:2509.19615},
  year={2025}
}

@article{jhaver2023decentralizing,
  title={Decentralizing platform power: A design space of multi-level governance in online social platforms},
  author={Jhaver, Shagun and Frey, Seth and Zhang, Amy X},
  journal={Social Media+ Society},
  volume={9},
  number={4},
  pages={20563051231207857},
  year={2023},
  publisher={SAGE Publications Sage UK: London, England}
}

@article{jia2024embedding,
  title={Embedding democratic values into social media AIs via societal objective functions},
  author={Jia, Chenyan and Lam, Michelle S and Mai, Minh Chau and Hancock, Jeffrey T and Bernstein, Michael S},
  journal={Proceedings of the ACM on Human-Computer Interaction},
  volume={8},
  number={CSCW1},
  pages={1--36},
  year={2024},
  publisher={ACM New York, NY, USA}
}

@article{kolluri2025alexandria,
  title={Alexandria: A Library of Pluralistic Values for Realtime Re-Ranking of Social Media Feeds},
  author={Kolluri, Akaash and Su, Renn and Jahanbakhsh, Farnaz and Zhao, Dora and Piccardi, Tiziano and Bernstein, Michael S},
  journal={arXiv preprint arXiv:2505.10839},
  year={2025}
}

@inproceedings{devito2017algorithms,
  title={" Algorithms ruin everything" \# RIPTwitter, Folk Theories, and Resistance to Algorithmic Change in Social Media},
  author={DeVito, Michael Ann and Gergle, Darren and Birnholtz, Jeremy},
  booktitle={Proceedings of the 2017 CHI conference on human factors in computing systems},
  pages={3163--3174},
  year={2017}
}

@inproceedings{smith2022recommender,
  title={Recommender systems and algorithmic hate},
  author={Smith, Jessie J and Jayne, Lucia and Burke, Robin},
  booktitle={Proceedings of the 16th ACM conference on recommender systems},
  pages={592--597},
  year={2022}
}

@article{wang2024trustworthy,
  title={Trustworthy recommender systems},
  author={Wang, Shoujin and Zhang, Xiuzhen and Wang, Yan and Ricci, Francesco},
  journal={ACM Transactions on Intelligent Systems and Technology},
  volume={15},
  number={4},
  pages={1--20},
  year={2024},
  publisher={ACM New York, NY}
}

@inproceedings{covington2016deep,
  title={Deep neural networks for youtube recommendations},
  author={Covington, Paul and Adams, Jay and Sargin, Emre},
  booktitle={Proceedings of the 10th ACM conference on recommender systems},
  pages={191--198},
  year={2016}
}

@article{ren2007applying,
  title={Applying common identity and bond theory to design of online communities},
  author={Ren, Yuqing and Kraut, Robert and Kiesler, Sara},
  journal={Organization studies},
  volume={28},
  number={3},
  pages={377--408},
  year={2007},
  publisher={Sage Publications Sage UK: London, England}
}

@article{fukuyama2020middleware,
  title={Middleware for dominant digital platforms: A technological solution to a threat to democracy},
  author={Fukuyama, Francis and Richman, Barak and Goel, Ashish and Katz, Roberta R and Melamed, A Douglas and Schaake, Marietje},
  journal={CyberPolicy Center, Freeman Spogli Institute},
  year={2020}
}

@article{lee2014social,
  title={Social media, network heterogeneity, and opinion polarization},
  author={Lee, Jae Kook and Choi, Jihyang and Kim, Cheonsoo and Kim, Yonghwan},
  journal={Journal of communication},
  volume={64},
  number={4},
  pages={702--722},
  year={2014},
  publisher={Oxford University Press}
}

@article{nechushtai2024more,
  title={More of the same? Homogenization in news recommendations when users search on Google, YouTube, Facebook, and Twitter},
  author={Nechushtai, Efrat and Zamith, Rodrigo and Lewis, Seth C},
  journal={Mass Communication and Society},
  volume={27},
  number={6},
  pages={1309--1335},
  year={2024},
  publisher={Taylor \& Francis}
}

@inproceedings{hsu2020awareness,
  title={Awareness, navigation, and use of feed control settings online},
  author={Hsu, Silas and Vaccaro, Kristen and Yue, Yin and Rickman, Aimee and Karahalios, Karrie},
  booktitle={Proceedings of the 2020 CHI Conference on Human Factors in Computing Systems},
  pages={1--13},
  year={2020}
}

@article{seering2022metaphors,
  title={Metaphors in moderation},
  author={Seering, Joseph and Kaufman, Geoff and Chancellor, Stevie},
  journal={New Media \& Society},
  volume={24},
  number={3},
  pages={621--640},
  year={2022},
  publisher={Sage Publications Sage UK: London, England}
}

@article{popowski2026social,
  title={Social Media Feed Elicitation},
  author={Popowski, Lindsay and Wu, Xiyuan and Zhu, Charlotte and Piccardi, Tiziano and Bernstein, Michael S},
  journal={arXiv preprint arXiv:2602.18594},
  year={2026}
}

@article{nogara2026longitudinal,
  title={A longitudinal analysis of misinformation, polarization and toxicity on Bluesky after its public launch},
  author={Nogara, Gianluca and Sahneh, Erfan Samieyan and DeVerna, Matthew R and Liu, Nick and Luceri, Luca and Menczer, Filippo and Pierri, Francesco and Giordano, Silvia},
  journal={Online Social Networks and Media},
  volume={51},
  pages={100342},
  year={2026},
  publisher={Elsevier}
}

@misc{custom-feeds-api,
key = {Custom Feeds.},
date = {2026},
key = {https://docs.bsky.app/docs/starter-templates/custom-feeds}
}

@misc{jaz-stats,
key = {Bluesky Post Count and Author Stats.},
date = {2026},
note = {https://bsky.jazco.dev/stats}
}

@misc{graze-marketplace,
key = {Graze Marketplace.},
date = {2026},
note = {https://www.graze.social/marketplace}
}

@misc{skyfeed,
    key = {SkyFeed.},
    note={https://skyfeed.app},
    date={2026}
}

@misc{graze,
    key = {Graze.},
    note={https://graze.social},
    date={2026}
}

@misc{blueskyfeedcreator,
    key = {Bluesky Feed Creator.},
    note={https://blueskyfeedcreator.com},
    date={2026}
}

@inproceedings{lampe2004slash,
  title={Slash (dot) and burn: distributed moderation in a large online conversation space},
  author={Lampe, Cliff and Resnick, Paul},
  booktitle={Proceedings of the SIGCHI conference on Human factors in computing systems},
  pages={543--550},
  year={2004}
}

@misc{attiebluesky,
    key={Meet Attie. Your research companion for the open social web.},
    note={https://attie.ai/},
    date={2026},
}

@inproceedings{bono2026self,
  title={Self-moderation in the decentralized era: decoding blocking behavior on bluesky},
  author={Bono, Carlo and Liu, Chang and Russo, Giuseppe and Pierri, Francesco},
  booktitle={Proceedings of the International AAAI Conference on Web and Social Media},
  volume={20},
  number={1},
  pages={315--329},
  year={2026}
}

@article{oshinowo2025seeing,
  title={Seeing the Politics of Decentralized Social Media Protocols},
  author={Oshinowo, Tolulope and Hwang, Sohyeon and Zhang, Amy X and Monroy-Hern{\'a}ndez, Andr{\'e}s},
  journal={arXiv preprint arXiv:2505.22962},
  year={2025}
}

@article{weld2025perceptions,
  title={Perceptions of Moderators as a Large-Scale Measure of Online Community Governance},
  author={Weld, Galen Cassebeer and Leibmann, Leon and Zhang, Amy X and Althoff, Tim},
  journal={Proceedings of the ACM on Human-Computer Interaction},
  volume={9},
  number={7},
  pages={1--29},
  year={2025},
  publisher={ACM New York, NY, USA}
}

@article{bernstein2023embedding,
  title={Embedding societal values into social media algorithms},
  author={Bernstein, Michael and Christin, Ang{\`e}le and Hancock, Jeffrey and Hashimoto, Tatsunori and Jia, Chenyan and Lam, Michelle and Meister, Nicole and Persily, Nathaniel and Piccardi, Tiziano and Saveski, Martin and others},
  journal={Journal of Online Trust and Safety},
  volume={2},
  number={1},
  year={2023}
}

@inproceedings{backstrom2016serving,
  title={Serving a billion personalized news feeds},
  author={Backstrom, Lars},
  booktitle={Proceedings of the Ninth ACM International Conference on Web Search and Data Mining},
  pages={469--469},
  year={2016}
}

@inproceedings{alvarado2018towards,
  title={Towards algorithmic experience: Initial efforts for social media contexts},
  author={Alvarado, Oscar and Waern, Annika},
  booktitle={Proceedings of the 2018 chi conference on human factors in computing systems},
  pages={1--12},
  year={2018}
}

@article{he2023cura,
  title={Cura: Curation at social media scale},
  author={He, Wanrong and Gordon, Mitchell L and Popowski, Lindsay and Bernstein, Michael S},
  journal={Proceedings of the ACM on Human-Computer Interaction},
  volume={7},
  number={CSCW2},
  pages={1--33},
  year={2023},
  publisher={ACM New York, NY, USA}
}

@article{failla2024m,
  title={“I’m in the Bluesky Tonight”: Insights from a year worth of social data},
  author={Failla, Andrea and Rossetti, Giulio},
  journal={PloS one},
  volume={19},
  number={11},
  pages={e0310330},
  year={2024},
  publisher={Public Library of Science San Francisco, CA USA}
}

@article{quelle2025bluesky,
  title={Bluesky: Network topology, polarization, and algorithmic curation},
  author={Quelle, Dorian and Bovet, Alexandre},
  journal={PloS one},
  volume={20},
  number={2},
  pages={e0318034},
  year={2025},
  publisher={Public Library of Science San Francisco, CA USA}
}

@article{keller2021future,
  title={The future of platform power: making middleware work},
  author={Keller, Daphne},
  journal={Journal of Democracy},
  volume={32},
  number={3},
  pages={168--172},
  year={2021},
  publisher={Johns Hopkins University Press}
}

@inproceedings{balduf2024looking,
  title={Looking at the blue skies of bluesky},
  author={Balduf, Leonhard and Sokoto, Saidu and Ascigil, Onur and Tyson, Gareth and Scheuermann, Bj{\"o}rn and Korczy{\'n}ski, Maciej and Castro, Ignacio and Kr{\'o}l, Micha{\l}},
  booktitle={Proceedings of the 2024 ACM on Internet Measurement Conference},
  pages={76--91},
  year={2024}
}

@article{ananny2018seeing,
  title={Seeing without knowing: Limitations of the transparency ideal and its application to algorithmic accountability},
  author={Ananny, Mike and Crawford, Kate},
  journal={new media \& society},
  volume={20},
  number={3},
  pages={973--989},
  year={2018},
  publisher={SAGE Publications Sage UK: London, England}
}

@article{swart2021experiencing,
  title={Experiencing algorithms: How young people understand, feel about, and engage with algorithmic news selection on social media},
  author={Swart, Jo{\"e}lle},
  journal={Social media+ society},
  volume={7},
  number={2},
  pages={20563051211008828},
  year={2021},
  publisher={SAGE Publications Sage UK: London, England}
}

@article{narayanan2023understanding,
  title={Understanding social media recommendation algorithms},
  author={Narayanan, Arvind},
  year={2023}
}

@inproceedings{klug2021trick,
  title={Trick and please. A mixed-method study on user assumptions about the TikTok algorithm},
  author={Klug, Daniel and Qin, Yiluo and Evans, Morgan and Kaufman, Geoff},
  booktitle={Proceedings of the 13th ACM web science conference 2021},
  pages={84--92},
  year={2021}
}

@inproceedings{raman2019challenges,
  title={Challenges in the decentralised web: The mastodon case},
  author={Raman, Aravindh and Joglekar, Sagar and Cristofaro, Emiliano De and Sastry, Nishanth and Tyson, Gareth},
  booktitle={Proceedings of the internet measurement conference},
  pages={217--229},
  year={2019}
}

@inproceedings{anaobi2023will,
  title={Will admins cope? Decentralized moderation in the fediverse},
  author={Anaobi, Ishaku Hassan and Raman, Aravindh and Castro, Ignacio and Zia, Haris Bin and Ibosiola, Damilola and Tyson, Gareth},
  booktitle={Proceedings of the ACM Web Conference 2023},
  pages={3109--3120},
  year={2023}
}

@article{hogg2024shaping,
  title={Shaping the Future of Social Media with Middleware},
  author={Hogg, Luke and DiResta, Ren{\'e}e and Fukuyama, Francis and Reisman, Richard and Keller, Daphne and Ovadya, Aviv and Thorburn, Luke and Stray, Jonathan and Mathur, Shubhi},
  journal={arXiv preprint arXiv:2412.10283},
  year={2024}
}

@article{braun2006using,
  title={Using thematic analysis in psychology},
  author={Braun, Virginia and Clarke, Victoria},
  journal={Qualitative research in psychology},
  volume={3},
  number={2},
  pages={77--101},
  year={2006},
  publisher={Taylor \& Francis}
}

@article{malki2025bonsai,
  title={Bonsai: Intentional and Personalized Social Media Feeds},
  author={Malki, Omar El and Qu{\'e}r{\'e}, Marianne Aubin Le and Monroy-Hern{\'a}ndez, Andr{\'e}s and Ribeiro, Manoel Horta},
  journal={arXiv preprint arXiv:2509.10776},
  year={2025}
}

@inproceedings{kleppmann2024bluesky,
  title={Bluesky and the at protocol: Usable decentralized social media},
  author={Kleppmann, Martin and Frazee, Paul and Gold, Jake and Graber, Jay and Holmgren, Daniel and Ivy, Devin and Johnson, Jeromy and Newbold, Bryan and Volpert, Jaz},
  booktitle={Proceedings of the ACM Conext-2024 Workshop on the Decentralization of the Internet},
  pages={1--7},
  year={2024}
}
